\documentclass{aa}  

\usepackage{graphicx,txfonts,lscape,natbib}
\bibpunct{(}{)}{;}{a}{}{,} % to follow the A&A style

\begin{document}
   \title{Metallicity, temperature, and gravity scales of M subdwarfs}

%   \subtitle{ }

   \author{N.\ Lodieu \inst{1,2}\thanks{Based on observations collected at the European Southern Observatory, Chile, under programmes 089.C-0140(A), 091.C-0264(A), 092.D-0600(A), and 093.C-0610(A)}, F.\ Allard \inst{3,4}, C.\ Rodrigo \inst{5,6}, Y.\ Pavlenko \inst{7,8,4}, A.\ Burgasser \inst{9}, Y.\ Lyubchik \inst{7}, B.\ Kaminsky \inst{7}, D.\ Homeier \inst{4,10}
          }

   \institute{Instituto de Astrof\'isica de Canarias (IAC), Calle V\'ia L\'actea s/n, E-38200 La Laguna, Tenerife, Spain. 
         \email{nlodieu@iac.es}
         \and
         Departamento de Astrof\'isica, Universidad de La Laguna (ULL), E-38205 La Laguna, Tenerife, Spain.
         \and
         Univ Lyon, Ens de Lyon, Univ Lyon1, CNRS, Centre de Recherche Astrophysique de Lyon UMR5574, F-69007, Lyon, France
         \and
         Visiting professor at the Instituto de Astrof\'isica de Canarias (IAC), La Laguna, Tenerife, Spain
         \and
         Departmento de Astrof\'isica, Centro de Astrobiolog\'ia (CSIC-INTA), ESAC Campus, Camino Bajo del Castillo s/n, E-286922, Villanueva de la Ca\~nada, Spain
         \and
         Spanish Virtual Observatory, E-28692 Villanueva de la Ca\~nada, Spain
         \and
         Main Astronomical Observatory of the National Academy of Sciences of Ukraine
         \and
         Center for Astrophysics Research, University of Hertfordshire, College Lane, Hatfield, Hertfordshire AL10 9AB, UK
         \and
         Center for Astrophysics and Space Science, University of California San Diego, La Jolla, CA 92093, USA
         \and
         Georg-August-Universit\"at Goettingen Institut f\"ur Astrophysik, Friedrich-Hund-Platz 1, 37077 G\"ottingen, Germany
             }

   \date{Received \today{}; accepted \today{}}
 
  \abstract
   {}
   {The aim of the project is to define a metallicity/gravity/temperature scale vs
spectral types for metal-poor M dwarfs.}
   {We obtained intermediate-resolution ultraviolet (R$\sim$3300), optical (R$\sim$5400), and near-infrared
(R$\sim$3900) spectra of 43 M subdwarfs (sdM), extreme subdwarfs (esdM), and ultra-subdwarfs (usdM)
with the X-shooter spectrograph on the European Southern Observatory Very Large Telescope. 
We compared our atlas of spectra to the latest BT-Settl synthetic spectral energy distribution over a 
wide range of metallicities, gravities, and effective temperatures to infer the physical 
properties for the whole M dwarf sequence (M0--M9.5) at sub-solar metallicities and constrain the latest 
state-of-the-art atmospheric models.
}
   {The BT-Settl models reproduce well the observed spectra across the 450--2500 nm
   wavelength range except for a few regions.
We find that the best fits are obtained for gravities of $\log$(g)\,=\,5.0--5.5 for the three 
metal classes. We infer metallicities of [Fe/H]\,=\,$-$0.5, $-$1.5, and $-$2.0$\pm$0.5 dex and 
effective temperatures of 3700--2600\,K, 3800--2900\,K, and 3700--2900\,K for subdwarfs, extreme subdwarfs, 
and ultra-subdwarfs, respectively. Metal-poor M dwarfs tend to be warmer by about 200$\pm$100\,K and
exhibit higher gravity than their solar-metallicity counterparts. We derive abundances of several
elements (Fe, Na, K, Ca, Ti) for our sample but cannot describe their atmospheres with a single metallicity 
parameter. Our metallicity scale expands the current scales available for midly metal-poor planet-host 
low-mass stars. Our compendium of moderate-resolution spectra covering the 0.45--2.5 micron range represents 
an important legacy value for large-scale surveys and space missions to come.
}
   {}


  \authorrunning{N.\ Lodieu}
  \titlerunning{Physical parameters of M subdwarfs}

   \maketitle
%
%________________________________________________________________

%
%%%%%%%%%%%%%%%%%%%%%%%%%%%%%%
%%%%%  Introduction  %%%%%
%%%%%%%%%%%%%%%%%%%%%%%%%%%%%%
%
\section{Introduction}
\label{SpClass_sdM:intro}

Mass, gravity, temperature, and metallicity constitute key parameters to understand the
formation and evolution of any type of stars. M dwarfs represent the largest community among members 
of the solar neighbourhood \citep{henry06,bochanski10,kirkpatrick12} and have also become attractive 
targets to search for Earth-like planets \citep[e.g.][]{nutzman08,quirrenbach12a,sozzetti14}. 
While the metal content of solar-type stars can be measured with great accuracies \citep[e.g.][]{adibekyan12a},
the metallicity of M dwarfs is more difficult to ascertain due to the significant amount of 
absorption bands \citep{kirkpatrick99}. Several groups conducted complementary surveys to assess 
the metallicity of M dwarfs based on photometry and colours 
\citep{bonfils05a,johnson09a,schlaufman10a,neves12,hejazi15,dittmann16}
and high-resolution spectra or spectral synthesis of multiple systems composed of a solar-type 
primary and a M dwarf secondary at optical and infrared wavelengths 
\citep{woolf05,woolf06,bean06a,woolf09,rojas_ayala10,rojas_ayala12,muirhead12,terrien12a,onehag12,neves13,neves14,hejazi15,newton15a,lindgren16a}.
These surveys usually focus on slightly metal-poor M dwarfs, with metallicities above $-$1.0 dex.
An extension to lower metallicities is needed to complement the precise distances that the Gaia 
satellite \citep{deBruijne12} will provide for a great number of spectroscopically-confirmed 
thick disk and halo M dwarfs \citep{savcheva14} but also understand the role of metallicity on 
the molecules and dust grains present in the atmosphere of low mass stars.

Subdwarfs are population II dwarfs which appear bluer than solar-metallicity stars 
due to the dearth of metals in their atmospheres \citep{baraffe97}.
They exhibit thick disk or halo kinematics, including high proper motions or large 
heliocentric velocities \citep{gizis97a}. They belong to the first generations of stars
and are important tracers of the chemical enrichment history of the Galaxy.
The original classification for M subdwarfs (sdM) and extreme subdwarfs (esdM)
developed by \citet{gizis97a} has been revised and extended by \citet{lepine07c}.
A new class of subdwarfs, the ultra-subdwarfs (usdM), has been added to the sdM 
and esdM. The new scheme is based on a parameter, $\zeta_{\rm TiO/CaH}$, which 
quantifies the weakening of the strength of the TiO band (in the optical) as a function 
of metallicity. Subdwarfs are easily distinguished from dwarfs with solar abundances 
because they exhibit stronger metal-hydride absorption bands (FeH, CrH) and metal 
lines (Ca{\small{I}}, Fe{\small{I}}) as well as blue infrared colours caused by
collision-induced H$_{2}$--H$_{2}$ absorption \citep{gizis97a,lepine07c,lodieu17a}. 
An independent classification scheme based on metallicity, gravity, and temperature 
has been proposed for M dwarfs by \citet{jao08} and an extension to the L dwarf regime
proposed by \citet{kirkpatrick14} and \citet{zhang17a}.

In this paper, we present moderate-resolution 0.45--2.5 $\mu$m spectroscopy of 
15 sdM, 16 esdM, and 12 usdM to infer their metallicities and T$_{\rm eff}$ by direct comparison 
with the latest BT-Settl synthetic models \citep{allard12}. 
In Section \ref{MplusTeff_sdM:sample} we present our sample of metal-poor M dwarfs 
drawn from the literature.
In Section \ref{MplusTeff_sdM:spec_obs} we describe our spectroscopic observations.
In Section \ref{MplusTeff_sdM:Models} we introduce the BT-Settl models used to infer the 
physical parameters of metal-poor M dwarfs.
In Section \ref{MplusTeff_sdM:Comparison} we infer the spectral type vs.\
metallicity/gravity/T$_{\rm eff}$ relation of the three M dwarf metal classes and
compare it to independent but complementary studies as well as solar-type M dwarfs.
In Section \ref{MplusTeff_sdM:discussion} we discuss our results and the
peculiarities of some of the spectra.

%
%%%%%%%%%%%%%%%%%%%%%%%%%%%%%%
%%%%% Sample selection %%%%%
%%%%%%%%%%%%%%%%%%%%%%%%%%%%%%
%
\section{Sample selection}
\label{MplusTeff_sdM:sample}

To select our sample of subdwarfs, we used the
SDSS spectroscopic database which contains a wealth of high-quality spectra covering
the 5000--9200\AA{} range at a spectral resolution of $\sim$2000\@. Hundreds of objects 
have been classified as sdM, esdM, or usdM following the scheme developed by \citet{lepine07c}.
This classification is publicly available through the SDSS archive and we have taken advantage 
of it to retrieve a large sample of low-metallicity M dwarfs.

We selected a sub-sample of low-metallicity stars which represent a sequence going from
M0 to M9.5 from optical spectra. The objects were specifically chosen to be the (or among the) 
brightest of their subclass and be observable either in August or January from
the Southern hemisphere to achieve the best signal-to-noise ratio possible over a wide
wavelength range. We completed our sample with LHS\,377 \citep[sdM7;][]{gizis97a},
observed with the same telescope/instrument but at higher spectral resolution 
\citep{rajpurohit16a}. Our final sample contains 15 sdM0--sdM9.5, 16 esdM0.0--esdM8.5, 
and 12 usdM0.0--usdM8.5) with almost one object per spectral sub-type. 
We are missing the sdM7.5, sdM8, and sdM9 in the subdwarf sequence, the esdM2.5 
and esdM8 in the extreme subdwarf sequence, and more usdM (usdM1.5, usdM2, usdM3.0, 
usdM6.5, usdM7, and usdM8). We note that these targets were included in our original
sample but not observed during the ESO service runs.
Table \ref{tab_MplusTeff_sdM:log_obs_Xshooter} lists their coordinates, optical SDSS$i$ 
magnitudes, optical spectral types on the \citet{lepine07c} scheme, dates of observations, 
mean airmass, exposure times in all three arms, and numbers of AB cycles for all 43 subdwarfs.

%
%%%%%%%%%%%%%%%%%%%%%%%%%%%%%%
%%%%% Observations %%%%%
%%%%%%%%%%%%%%%%%%%%%%%%%%%%%%
%
\section{VLT/X-shooter spectroscopy}
\label{MplusTeff_sdM:spec_obs}

We carried out spectroscopy from the UV- to the $K$ band with the X-Shooter
spectrograph \citep{dOdorico06,vernet11} mounted on the Cassegrain focus of the
Very Large Telescope (VLT) Unit 2\@. Observations were conducted in service
mode by the European Southern Observatory (ESO) staff over the course of 
three semesters, between August 2012 and June 2014 
(Table \ref{tab_MplusTeff_sdM:log_obs_Xshooter}). 
The conditions at the time of the observations met the clear sky request, with 
an airmass less than 1.6, grey conditions, and a seeing better than 1.2 arcsec. 
We set the individual on-source integration times according to the magnitudes
of the targets and used the multiple AB patterns to correct for the sky contribution 
(mainly) in the near-infrared. All observations were done with the slit oriented at 
parallactic angle. We list all the M subdwarfs observed with X-shooter in 
Table \ref{tab_MplusTeff_sdM:log_obs_Xshooter} along with a summary of the 
logs of the observations.

X-Shooter is a multi wavelength cross--dispersed echelle spectrograph made
of three arms covering simultaneously the ultraviolet (UVB; 0.3--0.55 $\mu$m), 
visible (VIS; 0.55--1.0 $\mu$m), and near--infrared (NIR; 1.0--2.48 $\mu$m) wavelength 
ranges thanks to the presence of two dichroics splitting the light. The spectrograph is 
equipped with three detectors: a 4096$\times$2048 E2V CCD44-82, a 4096$\times$2048 
MIT/LL CCID\,20, and a 2096$\times$2096 Hawaii 2RG for the UVB, VIS, and NIR arms,
respectively. We set the read-out mode to 400k and low gain without binning.
We used the 1.6 arcsec slit in the UVB, 1.5 arcsec in the VIS, and 1.2 arcsec
NIR, yielding resolving powers of 3300 (9.9 pixels per full-width-half-maximum), 
5400 (9.7 pixels per full-width-half-maximum), and 3900 (5.8 pixels per
full-width-half-maximum) in the UVB, VIS, and NIR arms, respectively.

We ran the latest version of the X-shooter pipeline (2.8.0)\footnote{http://www.eso.org/sci/software/pipelines/} 
on the raw data downloaded from the ESO archive with their associated raw calibration files 
from our three programmes: 089.C-0140(A), 091.C-0264(A), and 093.C-0610(A)\@. 
All spectra have their instrumental signature removed, including bias and flat-field. 
The spectra are wavelength-calibrated, sky-subtracted and finally flux-calibrated. 
The output products include a 2D spectrum associated with a 1D spectrum.
However, the optimal extraction of the 1D spectrum not being yet implemented in the version 2.8.0 
of the pipeline, we extracted the UVB, VIS, and NIR spectra with the {\tt{apsum}} 
task under IRAF \citep{tody86,tody93}.
We note that most of the targets have little flux in the UVB arm, resulting in low signal-to-noise
ratio below 550\,nm and very little flux (or no flux) below 400\,nm.
We corrected the VIS and NIR spectra for telluric bands/lines with the {\tt{molecfit}} package distributed by 
ESO \citep{kausch15,smette15}\footnote{http://www.eso.org/sci/software/pipelines/skytools/molecfit}
mainly because the telluric standards were not necessarily taken at the same airmass as our targets. 
The regions corrected are: 625--632, 686--696, 716--732, 758--770, 812--834, 893--920, 928--980\,nm in the 
VIS arm and 1105--1220, 1253--1280, 1310--1510, 1730--1995, 2000--2035, 2045--2085, 2200--2470\,nm in the NIR arm.

The final spectral energy distributions (SEDs) of sdM, esdM, and usdM are displayed in 
figures in Appendix \ref{MplusTeff_sdM:appendix_model_fits}. For display purposes, we shifted
the observed spectra to the BT-Settl models because of the large velocities of our subdwarfs 
taking into account that the wavelength scale of models is in vacuum and the observed
spectra in the air system (see Section \ref{MplusTeff_sdM:Comparison}). Our sample increases 
by a factor of five the sample of subdwarfs presented in \citet{rajpurohit16a} and extend that
recent work to the full sequence of the three metallicity classes of M subdwarfs.
We will make publicly available all spectra through the late-type subdwarf archive
at http://svo2.cab.inta-csic.es/vocats/ltsa/ \citep{lodieu17a}.

%
%%%%%%%%%%%%%%%%%%%%%%%%%%%%%%%%%%%%%%%%%
%%%%% Table: Log of observations: X-shooter %%%%%
%%%%%%%%%%%%%%%%%%%%%%%%%%%%%%%%%%%%%%%%%
%
\begin{table*}
 \centering
 \caption[]{Logs of the VLT/X-shooter spectroscopic observations of M subdwarfs.
We list the coordinates and Sloan $i'$ magnitude, the spectral type from
by the SDSS spectroscopic database, the date of the VLT X-shooter with the
airmass at the beginning of the observations, the on-source integrations in
each arm (UVB/VIS/NIR) with the numbers of AB cycles (identical for all arms).
}
 \begin{tabular}{@{\hspace{0mm}}c @{\hspace{1mm}}c c c c c c c@{\hspace{0mm}}}
 \hline
 \hline
R.A\ (J2000)  & dec (J2000) & $i'$ & SpT & Date     & Airm & ExpT           & Cycles \cr
 \hline
hh:mm:ss.ss   & dd:mm:ss.s  & mag &    &       &         & sec/sec/sec   &        \cr
 \hline
23:46:59.87 & $-$00:59:43.9 & 15.166 & sdM0.0 & 13Aug2012 & 1.30 &  90/66/66   & 2AB \cr
21:22:02.76 & $+$00:44:56.8 & 14.996 & sdM0.5 & 21May2012 & 1.13 &  90/66/66   & 2AB \cr
00:48:00.05 & $+$00:28:49.4 & 15.260 & sdM1.0 & 13Aug2012 & 1.22 &  90/66/66   & 2AB \cr
00:51:25.68 & $-$00:37:16.9 & 16.666 & sdM1.5 & 18Aug2012 & 1.10 & 216/216/240 & 2AB \cr
23:57:32.54 & $-$01:10:36.3 & 15.692 & sdM2.0 & 17Aug2012 & 1.32 &  90/66/66   & 2AB \cr
01:17:46.53 & $-$00:01:05.3 & 16.196 & sdM2.5 & 13Aug2012 & 1.36 &  90/66/66   & 2AB \cr
03:24:49.81 & $-$00:15:05.0 & 14.942 & sdM3.0 & 15Aug2012 & 1.14 &  90/66/66   & 2AB \cr
02:12:08.59 & $+$00:37:01.5 & 14.918 & sdM3.5 & 19Aug2012 & 1.11 &  90/66/66   & 2AB \cr
03:46:01.58 & $+$00:55:11.6 & 15.872 & sdM4.0 & 19Aug2012 & 1.14 &  90/66/66   & 2AB \cr
22:57:48.05 & $+$14:29:39.9 & 18.336 & sdM4.5 & 07Jul2013 & 1.39 & 276/292/300 & 5AB \cr
00:48:36.45 & $+$00:09:31.7 & 17.175 & sdM5.0 & 10Jan2012 & 1.32 &  90/66/66   &  AB \cr
23:53:36.94 & $+$00:47:34.1 & 18.025 & sdM5.5 & 14Jul2013 & 1.26 & 276/292/300 & 5AB \cr
00:25:52.59 & $+$01:09:24.9 & 17.797 & sdM6.0 & 19Aug2012 & 1.18 & 266/266/290 & 5AB \cr
16:10:28.99 & $-$00:40:53.0 & 15.903 & sdM6.5 & 10Jun2012 & 1.11 &  90/66/66   & 2AB \cr
02:05:33.75 & $+$12:38:24.0 & 18.107 & sdM8.0 & 11Aug2012 & 1.30 & 266/266/290 & 5AB \cr
10:13:07.35 & $-$13:56:20.4 & 16.010 & sdM9.5 & 09May2013 & 1.34 & 126/142/150 & 1AB \cr
\hline
01:37:55.30 & $-$09:39:41.9 & 17.488 & esdM0.0 & 08Sep2013 & 1.45 & 276/292/300 & 5AB \cr
23:25:41.30 & $+$00:04:19.6 & 16.203 & esdM0.5 & 18Aug2012 & 1.13 &  90/66/66   & 2AB \cr
00:38:02.86 & $+$00:50:14.2 & 17.520 & esdM1.0 & 14Jul2013 & 1.20 & 276/292/300 & 5AB \cr
23:40:39.45 & $-$00:51:18.4 & 16.263 & esdM1.5 & 12Aug2012 & 1.49 &  90/66/66   & 2AB \cr
12:55:29.19 & $-$03:30:55.8 & 17.536 & esdM2.0 & 11Jul2013 & 1.18 & 276/292/300 & 5AB \cr
12:53:53.35 & $+$12:12:48.7 & 16.531 & esdM3.0 & 11Jul2013 & 1.30 & 126/142/150 & 3AB \cr
14:52:04.63 & $+$10:18:26.3 & 17.892 & esdM3.5 & 30Jun2014 & 1.22 & 276/292/300 & 5AB \cr
00:40:18.18 & $-$10:41:55.9 & 17.405 & esdM4.0 & 18Aug2012 & 1.04 & 276/292/300 & 5AB \cr
01:09:54.11 & $-$10:12:12.6 & 14.766 & esdM4.5 & 04Aug2013 & 1.05 & 126/142/150 & 1AB \cr
13:31:51.15 & $-$00:09:19.0 & 18.180 & esdM5.0 & 15Jun2013 & 1.16 & 276/292/300 & 5AB \cr
09:03:07.95 & $+$08:42:43.2 & 15.988 & esdM5.5 & 18Apr2014 & 1.20 & 126/142/150 & 1AB \cr
15:36:47.08 & $+$02:55:01.5 & 19.071 & esdM6.0 & 12Aug2012 & 1.14 & 276/276/300 & 6AB \cr
01:33:46.24 & $+$13:28:22.4 & 17.830 & esdM6.5 & 04Aug2013 & 1.37 & 276/292/300 & 5AB \cr
02:35:57.61 & $+$01:08:00.5 & 19.190 & esdM7.0 & 09Sep2013 & 1.11 & 456/472/480 & 4AB \cr
05:58:58.91 & $-$29:03:26.7 & 16.320 & esdM7.5 & 27Aug2012 & 1.34 & 216/216/240 & 2AB \cr
04:52:09.94 & $-$22:45:08.4 & 17.220 & esdM8.5 & 17Aug2012 & 1.13 & 216/216/240 & 2AB \cr
\hline
03:27:28.10 & $-$00:50:01.4 & 17.462 & usdM0.0 & 16Spe2012 & 1.09 & 266/266/290 & 5AB \cr
15:34:04.63 & $+$09:36:22.5 & 17.984 & usdM0.5 & 09Aug2013 & 1.22 & 266/266/290 & 5AB \cr
15:12:18.36 & $+$09:30:40.7 & 17.890 & usdM1.0 & 23Jul2013 & 1.23 & 276/292/300 & 5AB \cr
20:59:20.57 & $+$00:00:33.4 & 17.853 & usdM2.5 & 11Aug2012 & 1.12 & 266/266/290 & 5AB \cr
10:41:07.20 & $+$06:33:04.7 & 18.062 & usdM3.0 & 31May2013 & 1.27 & 246/262/270 & 5AB \cr
15:35:40.74 & $+$08:21:43.3 & 18.008 & usdM4.0 & 10Aug2013 & 1.21 & 276/292/300 & 5AB \cr
14:17:48.62 & $+$07:11:05.4 & 17.658 & usdM4.5 & 12Jul2013 & 1.18 & 276/292/300 & 5AB \cr
12:04:26.91 & $+$13:29:23.3 & 16.549 & usdM5.0 & 12Jun2012 & 1.33 & 216/216/240 & 2AB \cr
16:27:54.22 & $+$00:37:14.0 & 18.384 & usdM5.5 & 17Aug2012 & 1.15 & 266/266/290 & 5AB \cr
16:41:23.73 & $+$24:49:42.4 & 17.233 & usdM6.0 & 15Jul2013 & 1.58 & 126/142/150 & 3AB \cr
08:22:33.69 & $+$17:00:19.9 & 17.300 & usdM7.5 & 03Apr2013 & 1.39 & 126/142/150 & 3AB \cr
12:27:05.06 & $-$04:47:20.7 & 16.630 & usdM8.5 & 11Jul2013 & 1.35 & 126/142/150 & 3AB \cr
 \hline
 \label{tab_MplusTeff_sdM:log_obs_Xshooter}
 \end{tabular}
\end{table*}
%

%
%%%%%%%%%%%%%%%%%%%%%%%%%%%%%%%%%%%%%%
%%%%%  Models  %%%%%
%%%%%%%%%%%%%%%%%%%%%%%%%%%%%%%%%%%%%%
%
\section{BT-Settl synthetic spectra}
\label{MplusTeff_sdM:Models}

To infer the range of physical parameters (gravities i.e.\ $\log$(g), metallicities [M/H], and T$_{\rm eff}$)
for our sequence of subdwarfs, we employed the BT-Settl models 
\citep{allard03a,allard07a,allard10a,allard11,allard12} available for retrieval at France Allard's 
webpage\footnote{http://perso.ens-lyon.fr/france.allard/}. The BT-Settl models account for TiO 
\citep{plez98,plez08} and H$_{2}$O (\citep{barber06a} among other opacities using the \citet{caffau11a}
abundance values and mixing information for the CO$^{5}$BOLD code \citep{steiner07,freytag10a}.
These models are valid for a wide range of T$_{\rm eff}$ (400--8000\,K), gravities 
($\log$(g)\,=\,2.5--6.0 dex), and metallicities (from $-$5.0 to solar).
The synthetic SEDs span the wavelength range from 10\AA{} up to 1000\,$\mu$m.
The stellar metallicity is defined by the total iron content of a star because iron is the easiest 
species to measure spectroscopically. The abundance ratio [Fe/H] is defined as the logarithm of the 
ratio of a star's iron abundance compared to that of the Sun, where $-$1.0 dex means that a star has 
one-tenth of the solar metallicity. The determination of the metallicity is only valid if the abundance
of all elements follow the abundance of iron. However, this statement might not be totally true in the
case of halo dwarfs that suffered different nucleosynthesis events.

We downloaded the spectra from the CIFIST2011 grid\footnote{phoenix.ens-lyon.fr/Grids/BT-Settl/CIFIST2011/SPECTRA/}
and considered the ranges that encompass the spectral types of the three metallicity classes:
T$_{\rm eff}$\,=\,4000--2500\,K, gravity ($\log$g\,=\,4.5--5.5 dex), and metallicity [M/H] 
(from $-$2.5 dex to solar).
These models assumes solar abundance values (Z\,=\,0.0153, Z/X\,=\,0.0209) from \citet{caffau11a}
with alpha enhancement taken into account as follows: [alpha/H]=$+$0.2 relative to solar for [M/H]\,=\,$-$0.5 
dex and [alpha/H]\,=\,$+$0.4 for lower metallicities.
For direct comparison, we limited the wavelength range to 450--2500 nm and
smoothed the synthetic SEDs with the Interactive Data Language (IDL) {\tt{gaussfold}} 
function\footnote{http://astro.uni-tuebingen.de/software/idl/aitlib/misc/gaussfold.html} to the 
spectral resolution of our X-shooter spectra.

Throughout the paper, we use the term metallicity [Fe/H] which is given by the BT-Settl models.
We do not infer abundances or metallicities of single element but the global values given by
the synthetic spectra. The former can differ from the latter as shown for a metal-poor low-mass
binary \citep{pavlenko15}. The study of abundances of individual elements in subdwarfs is beyond
the scope of this paper but will be investigated in a future publication.

%
%%%%%%%%%%%%%%%%%%%%%%%%%%%%%%%%%%%%%%
%%%%%  Comparison  %%%%%
%%%%%%%%%%%%%%%%%%%%%%%%%%%%%%%%%%%%%%
%
\section{Comparison: observations vs.\ models}
\label{MplusTeff_sdM:Comparison}

We fitted the observed spectra with the BT-Settl SEDs with a chi-square ($\chi^2$) minimisation procedure
described in Sect.\ \ref{MplusTeff_sdM:Comparison_chi2}. We derived the temperature 
(Sect.\ \ref{MplusTeff_sdM:Comparison_Teff}) and metallicity (Sect.\ \ref{MplusTeff_sdM:Comparison_MH})
scales from the best model fits.

\subsection{Chi-square fitting}
\label{MplusTeff_sdM:Comparison_chi2}

We performed a $\chi^2$ fit to compare the observed spectra with those in the BT-Settl
theoretical library. We considered the following ranges in temperature, gravity, and metallicity:
4000--2500\,K, $\log$g\,=\,4.5--6.0 dex, and [M/H] between $-$3.0 and 0.0 dex, as expected for
old low-mass M-type dwarfs. The steps are 100\,K and 0.5 dex in temperatures and gravity $+$ metallicity,
respectively. We ignored regions of the observed spectra strongly affected by telluric bands, 
in particular the 530--570 nm, 928--1010 nm, 1110--1150 nm, 1340--1460 nm, 1790--1970 nm wavelength ranges.

We shifted our observed spectra to the wavelength of the models. We calculated the heliocentric 
radial velocities for our sample of metal-poor M dwarfs using a set of about 15 lines
(potassium, sodium, iron, and calcium) in the
610--840 nm wavelength range, the exact number depending on the quality of the spectrum and
strength of the line (Table \ref{tab_MplusTeff_sdM:table_param}). We measured
consistent shifts between these five strong lines for all sources, with dispersions of 
the order of a few km/s. We assume that the true error bars are set 
by the resolution of our X-shooter spectra (3900--6700), corresponding to radial velocity
uncertainties of approximately 15--25 km/s.

Then, we minimized the $\chi^2$ value for each observed and theoretical spectrum, as:

\begin{equation}\label{eq:chi2}
\chi^2 = \frac{1}{N} \sum_i\left\{ \frac{ ({\rm F}_{\rm obs,i} - A \ {\rm F}_{\rm mod,i})^2}{(\Delta{\rm F}_{\rm obs,i})^2}\right\}
\end{equation}

\noindent The scale factor $A$ is calculated to minimize $\chi^2$ for each case as:

\begin{equation}\label{eq:chi2b}
A = \frac{ \sum_i({\rm F}_{\rm obs,i} {\rm F}_{\rm mod,i}/ \Delta{\rm F}_{\rm obs,i}^2)}{\sum_i({\rm F}_{\rm mod,i}^2/\Delta{\rm F}_{\rm obs,i}^2)}
\end{equation}

\noindent In both expressions the sum is performed over the full wavelength range. ${\rm F}_{\rm obs,i}$ 
and $\Delta{\rm F}_{\rm obs,i}$ are the observed values of the flux and its associated errors, respectively.
${\rm F}_{\rm mod,i}$ are the corresponding values of the theoretical spectrum.

To avoid the fact that points with small observational errors have an excessive weight in the fitting process, 
we calculated the average of the error values in each spectrum as:

\begin{equation}\label{eq:chi2c}
|\Delta{\rm F}_{\rm obs}| = \frac{1}{N} \sum_i\Delta{\rm F}_{\rm obs,i}
\end{equation}

\noindent Then, we fixed the error to be half the average for those points with smaller errors during the
fitting process.

Overall, we find that the BT-Settl reproduce well the SEDs of metal-poor M dwarfs and in particular
the main molecular bands. However, we noticed that some of the main atomic lines are not well
reproduced: ((Figs.\ \ref{fig_MplusTeff_sdM:full_XSH_model_lines_sdM},
\ref{fig_MplusTeff_esdM:full_XSH_model_lines_esdM}, \ref{fig_MplusTeff_usdM:full_XSH_model_lines_usdM}).
The lines of the sodium doublet at $\sim$820 nm predicted by the BT-Settl models appear too broad for 
spectral types later than approx M5\@. The calcium lines seem too narrow for ultra-subdwarfs but 
look correctly reproduced for subdwarfs and extreme subdwarfs. This mis-match between observations
and models yields over or under estimates of the abundances of these elements. The disappearance of 
these elements in other molecules (e.g.\ CaOH) could also explain the observed discrepancy.
We note that the potassium doublet at around 760/790nm is well reproduced by the models, except
for the latest spectral types ($\geq$M7) and the coolest sources
(Figs.\ \ref{fig_MplusTeff_sdM:full_XSH_model_lines_sdM}, \ref{fig_MplusTeff_esdM:full_XSH_model_lines_esdM},
\ref{fig_MplusTeff_usdM:full_XSH_model_lines_usdM}).

As a consequence, we opted for four fitting procedures, described below, to gauge the uncertainties 
on the physical parameters derived from the synthetic spectra. We describe them below:
\begin{enumerate}
\item ``FF'' corresponds to the fit of the full SED of each subtype and metal class from 450 to 2500 nm
(Figs.\ \ref{fig_MplusTeff_sdM:full_XSH_model_FF_sdM}--\ref{fig_MplusTeff_sdM:full_XSH_model_FF_usdM}).
\item ``LL'' corresponds the fitting procedure of a few lines in the optical spectra (K{\small{I}} and 
Na{\small{I}}). The model spectra shown correspond to the physical parameters derived from the line fits 
(Figs.\ \ref{fig_MplusTeff_sdM:full_XSH_model_LL_sdM}--\ref{fig_MplusTeff_sdM:full_XSH_model_LL_usdM}).
\item ``FL'': in this case we fix the temperature derived from the full fit and adjust gravity and
metallicity to converge towards the best fit of the aforementioned lines
(Figs.\ \ref{fig_MplusTeff_sdM:full_XSH_model_FL_sdM}--\ref{fig_MplusTeff_sdM:full_XSH_model_FL_usdM}).
\item ``LF'': we fix the gravity and metallicity from the fits of the line and adjust the effective
temperature fitting the full SED of each subtype and subclass.
(Figs.\ \ref{fig_MplusTeff_sdM:full_XSH_model_LF_sdM}--\ref{fig_MplusTeff_sdM:full_XSH_model_LF_usdM}).
\end{enumerate}

In general, we find that the ``FF'' fitting procedure reproduces best all observed spectra. Therefore,
we conclude that the physical parameters derived from this option are the most probable.

We list the model-dependent physical parameters in Table \ref{tab_MplusTeff_sdM:table_param} and display 
the best fits provided by the synthetic SEDs (red lines) to observed X-shooter spectra (black lines) 
for each metal class in Appendix \ref{MplusTeff_sdM:appendix_model_fits}.

We have also looked at the physical parameters derived from the model fit to the optical region of
the X-shooter spectra (600--1000nm) from which the spectral classification of subdwarfs is based
\citep{gizis97a,lepine07c}.
We find that on average the optical spectra give equal or cooler effective temperatures, lower gravites,
and/or lower metallicities
(Table \ref{tab_MplusTeff_sdM:table_best_fits_VV} in Appendix \ref{MplusTeff_sdM:appendix_model_fits}).

%
%%%%%%%%%%%%%%%%%%%%%%%%%%%%%%%%%%%%%%%
%%%%% Table: Physical parameters subdwarfs %%%%%
%%%%%%%%%%%%%%%%%%%%%%%%%%%%%%%%%%%%%%%
%
\begin{table*}
 \centering
 \caption[]{
Adopted physical parameters for sdM, esdM, and usdM from the comparison between the observed VLT/X-shooter 
spectra and the BT-Settl synthetic spectra from the fit of the overall SED\@. For each spectral subtype 
and metal class, we list the radial velocities (RV in km/s), gravity ($\log$g), temperature 
(T$_{\rm eff}$ in K), and metallicity ([Fe/H]). Uncertainties on the radial velocities, gravities, 
T$_{\rm eff}$, and metallicities are 15--25 km/s, 0.5 dex, 100\,K, and 0.25 dex, respectively. 
Uncertainties on mass from the \citet{baraffe97} models is approximately $\pm$10\%.
No X-shooter spectrum is available for the subtype with missing values.
$^{a}$ LHS\,377 parameters are from \citet{rajpurohit16a}.
}
 \begin{tabular}{@{\hspace{0mm}}c @{\hspace{1mm}}c @{\hspace{1mm}}c c c c c c c c c c c c c c c@{\hspace{0mm}}}
 \hline
 \hline
  &   &  \multicolumn{4}{c}{sdM} & \multicolumn{5}{c}{esdM} & \multicolumn{5}{c}{usdM} \cr
 \hline
SpT    & RV & $\log$g & T$_{\rm eff}$ & [Fe/H] & Mass & RV & $\log$g & T$_{\rm eff}$ & [Fe/H] & Mass & RV & $\log$g & T$_{\rm eff}$ & [Fe/H] & Mass \cr
 \hline
       & km// & dex     &      K        &  dex  & M$_{\odot}$ & km/s & dex     &      K        &  dex & M$_{\odot}$  & km/s & dex     &      K        &  dex & M$_{\odot}$  \cr
 \hline
M0.0   & $+$122.6 & 4.5  & 3600 & $-$1.5 & 0.133 & $-$9.2   & 5.0  & 3800 & $-$1.0 & 0.275 & $+$122.6 & 4.5  & 3500 & $-$2.0 & 0.106 \cr
M0.5   & $-$45.1  & 5.0  & 3700 & $-$1.0 & 0.216 & $-$97.3  & 5.0  & 3700 & $-$1.5 & 0.153 & $-$117.1 & 5.5  & 3700 & $-$2.5 & 0.125 \cr
M1.0   & $-$33.7  & 5.0  & 3700 & $-$0.5 & 0.352 & $-$259.7 & 5.5  & 3800 & $-$0.5 & 0.421 & $-$15.9  & 5.5  & 3700 & $-$1.0 & 0.216 \cr
M1.5   & $-$78.7  & 4.5  & 3600 & $-$0.0 & 0.600 & $-$46.7  & 5.0  & 3600 & $-$1.0 & 0.178 &    ---   & ---  &  --- &   ---  &  ---  \cr
M2.0   & $-$133.8 & 5.5  & 3600 & $-$0.5 & 0.272 & $+$40.6  & 5.5  & 3600 & $-$1.0 & 0.178 &    ---   & ---  &  --- &   ---  &  ---  \cr
M2.5   & $-$39.7  & 5.5  & 3600 & $-$0.5 & 0.272 &    ---   & ---  &  --- &   ---  &  ---  & $-$138.8 & 5.5  & 3600 & $-$1.5 & 0.133 \cr
M3.0   & $-$52.6  & 5.0  & 3500 & $-$0.0 & 0.450 & $+$35.3  & 5.5  & 3500 & $-$1.5 & 0.120 & $+$178.0 & 6.0  & 3800 & $-$0.5 & 0.421 \cr
M3.5   & $-$21.7  & 5.0  & 3400 & $-$0.5 & 0.177 & $-$52.5  & 5.5  & 3500 & $-$1.5 & 0.120 &    ---   & ---  &  --- &   ---  &  ---  \cr
M4.0   & $-$67.2  & 5.0  & 3400 & $-$1.0 & 0.129 & $-$81.3  & 5.5  & 3400 & $-$1.5 & 0.110 & $-$19.8  & 4.5  & 3200 & $-$1.5 & 0.098 \cr
M4.5   & $-$470.7 & 5.5  & 3300 & $-$1.0 & 0.117 & $-$60.7  & 5.5  & 3400 & $-$1.0 & 0.129 & $-$102.5 & 5.5  & 3400 & $-$2.0 & 0.100 \cr
M5.0   & $-$31.6  & 5.0  & 3200 & $-$0.0 & 0.150 & $+$21.7  & 4.5  & 3200 & $-$1.5 & 0.098 & $+$180.0 & 5.5  & 3500 & $-$2.5 & 0.106 \cr
M5.5   & $-$12.5  & 5.5  & 3200 & $-$1.0 & 0.107 & $+$318.1 & 5.5  & 3300 & $-$2.0 & 0.098 & $-$25.1  & 4.5  & 3300 & $-$2.0 & 0.098 \cr
M6.0   & $-$87.8  & 5.5  & 3200 & $-$1.0 & 0.107 & $-$38.5  & 5.5  & 3300 & $-$2.0 & 0.098 & $+$104.8 & 5.5  & 3300 & $-$2.0 & 0.098 \cr
M6.5   & $-$57.2  & 5.5  & 2900 & $-$0.0 & 0.063 & $-$3.7   & 5.5  & 3300 & $-$2.0 & 0.098 &    ---   & ---  &  --- &   ---  &  ---  \cr
M7.0   &    ---   & 5.0  & 3000 & $-$1.0$^{a}$ & 0.096 & $-$167.1 & 5.5  & 3200 & $-$0.0 & 0.096 &    ---   & ---  &  --- &   ---  &  ---  \cr
M7.5   &    ---   & ---  &  --- &  ---  &  ---  & $+$191.1 & 5.0  & 3000 & $-$2.0 & 0.090 & $+$122.1 & 5.5  & 3100 & $-$2.5 & 0.092 \cr
M8.0   & $-$208.9 & 5.5  & 2900 & $-$2.0 & 0.089 &   ---    & ---  &  --- &   ---  &  ---  &    ---   & ---  &  --- &   ---  &  ---  \cr
M8.5   &   ---   & ---  &  --- &  ---  &  ---   & $+$72.7  & 5.5  & 3000 & $-$2.0 & 0.090 & $+$85.0  & 5.5  & 3100 & $-$2.5 & 0.092 \cr
M9.5   & $+$64.7  & 5.0  & 2800 & $-$2.0 & 0.088 &   ---    & ---  &  --- &   ---  &  ---  &    ---   & ---  &  --- &   ---  &  ---  \cr
 \hline
 \label{tab_MplusTeff_sdM:table_param}
 \end{tabular}
\end{table*}
\subsection{Temperature scale}
\label{MplusTeff_sdM:Comparison_Teff}

\begin{figure}
  \centering
  \includegraphics[width=\linewidth, angle=0]{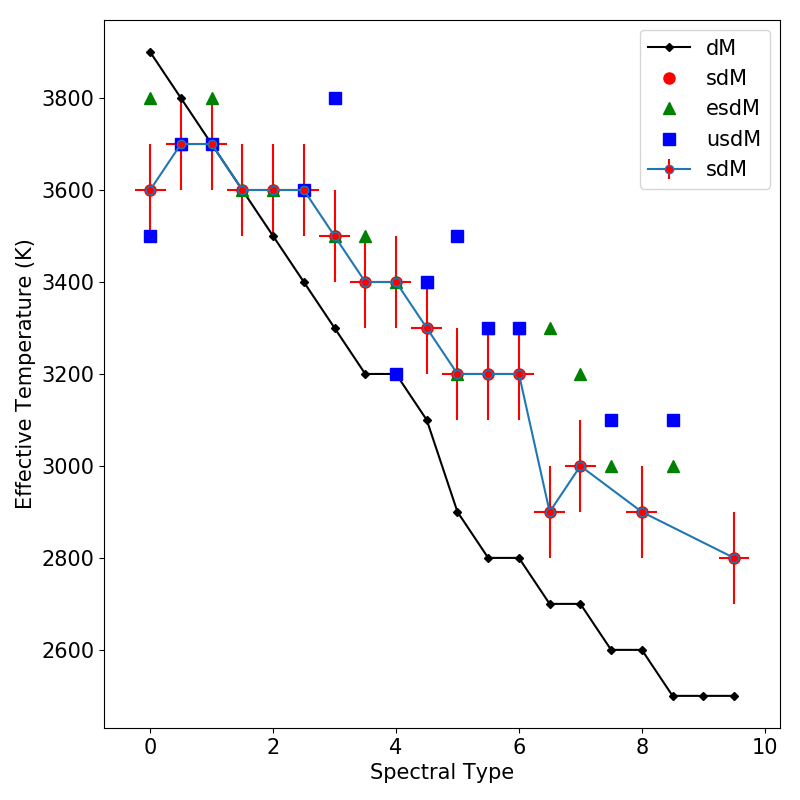}
   \caption{T$_{\rm eff}$ as a function of spectral type for solar-metallicity M dwarfs 
   (black diamonds with line), sdM (red dots), esdM (green triangles) and usdM (blue squares).
   Error bars on spectral types and model parameters are marked for sdM only for clarity purposes.
   Error bars on spectral type and temperatures are $\pm$0.25 and $\pm$100\,K, respectively.
   The point for the sdM7 subdwarf corresponds to LHS\,377 from  \citet{rajpurohit16a}.
}
   \label{fig_MplusTeff_sdM:plot_SpT_Teff}
\end{figure}

We derived comparable T$_{\rm eff}$ intervals for all three metallicity classes 
(Table \ref{tab_MplusTeff_sdM:table_param}) fitting the full SEDs 
(Tables \ref{tab_MplusTeff_sdM:table_best_fits_FF} and \ref{tab_MplusTeff_sdM:table_best_fits_FL}).
The T$_{\rm eff}$ range from $\sim$3800\,K for the earlier M subdwarfs down to $\sim$2600\,K for the
latest spectral types. We can hardly distinguish the three classes in the T$_{\rm eff}$ vs.\ spectral 
type diagram shown in Fig.\ \ref{fig_MplusTeff_sdM:plot_SpT_Teff} within the error bars of 100\,K set
by the steps available in the models.
We overplotted the temperature scale of field M dwarfs (solid black line in 
Fig.\ \ref{fig_MplusTeff_sdM:plot_SpT_Teff})from the latest relation of \citet{rajpurohit13}, whose 
trend is comparable to earlier studies within error bars \citep{bessell91,leggett96,leggett00a,testi09a}.
Dwarfs with spectral types earlier than M2 are indistinguishable in the temperature parameter space.
Overall, the temperatures of metal-poor M dwarfs are similar to those of solar-type M dwarfs
with an offset of 200$\pm$100\,K towards warmer temperatures.
They follow a linear trend with some spectral types being off by 100\,K or 200\,K, which may be due to
the error on the spectral classification or binarity which is mainly based on a spectral index measuring 
the strength of the CaH and TiO bands in the optical \citep{lepine07c}.

We note that the temperature scale of the line fitting option tends to infer lower effective
temperatures by at least 100\,K with a similar interval for temperatures hotter than 3400\,K, 
producing ranges of 3700--2700\,K. The agreement is better at lower temperatures and within
the uncertainty of 100\,K
(Tables \ref{tab_MplusTeff_sdM:table_best_fits_LL} and \ref{tab_MplusTeff_sdM:table_best_fits_LF}).
We also find that the fit to the optical region only yields typically lower effective temperature by 100\,K to 200\,K 
(Table \ref{tab_MplusTeff_sdM:table_best_fits_VV} in Appendix \ref{MplusTeff_sdM:appendix_model_fits}).

%
%%%%%%%%%%%%%%%%%%%%%%%%%%%%%%%%%%%%%%
%%%%% Figure: SpType vs Teff %%%%%
%%%%%%%%%%%%%%%%%%%%%%%%%%%%%%%%%%%%%%
%
\subsection{Gravity scale}
\label{MplusTeff_sdM:Comparison_grav}

Subdwarfs are old low-mass stars that belong to the thick diskor halo of our Galaxy. On average,
they are much older than their solar-metallicity counterparts. \citet{monteiro06a} identified
several white dwarf-subdwarf systems in the thick disk and derived ages of 6--9 Gyr for two of 
them with M subdwarf companions. From the comparison with FGK stellar templates selected as
benchmarks for the Gaia mission \citep{jofre14a}, \citet{scholz15b} inferred a possible age of 
12 Gyr for an ancient metal-poor ($-$2.0$\pm$0.2 dex) F-type star member of the Galactic halo
due to its large tangential velocity. On average, we expect our sample to be older than 5 Gyr.

\begin{figure}
  \centering
  \includegraphics[width=\linewidth, angle=0]{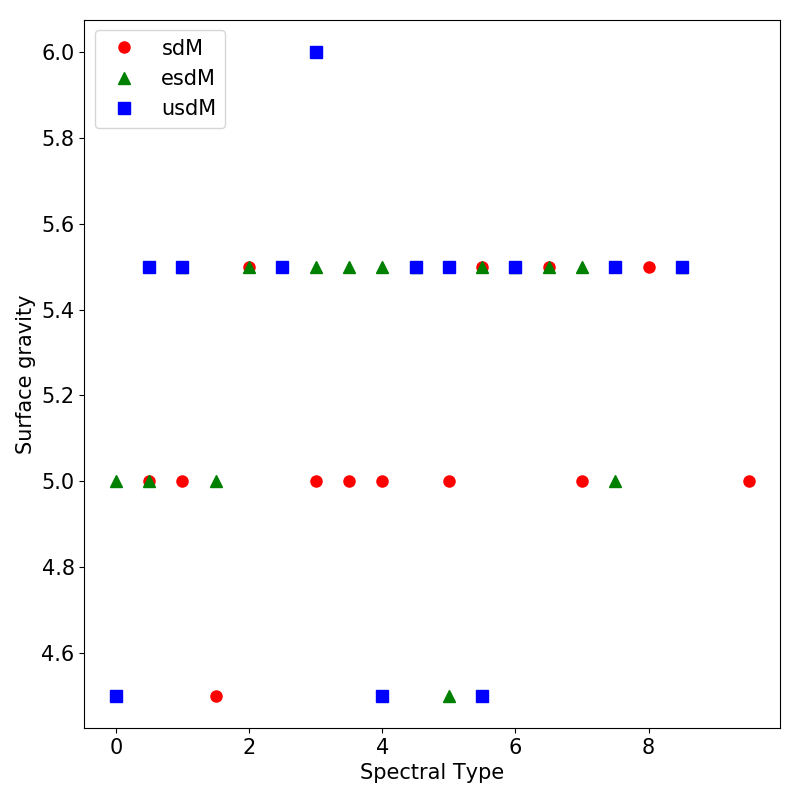}
   \caption{Gravity as a function of spectral type for sdM (red dots), esdM (green triangles)
   and usdM (blue squares). Error bars on spectral type and gravity are $\pm$0.5
   and $\pm$0.5 dex, respectively.
   The point for the sdM7 subdwarf corresponds to LHS\,377 from  \citet{rajpurohit16a}.
}
   \label{fig_MplusTeff_sdM:plot_SpT_logg}
\end{figure}

From the fit of the  BT-Settl models to the full VIS$+$NIR spectral energy distribution of our
M subdwarfs, we infer gravities of 4.5--5.5 dex for all metal classes, except for one object 
with $\log$(g)\,=\,6.0 dex (usdM3.0, see discussion section).
However, we observe a possible trend of increasing mean gravity with lower metallicity, going from 
5.0 dex for sdM to 5.5 dex for esdM and usdM (Fig.\ \ref{fig_MplusTeff_sdM:plot_SpT_logg}).
We note that the range of gravities of the CIFIST models is limited to 6.0 dex, higher gravities
are desirable to corroborate this statement.
We also note that the fit to the optical region only yields typically lower gravities by 0.5 to 1.0 dex 
(Table \ref{tab_MplusTeff_sdM:table_best_fits_VV} in Appendix \ref{MplusTeff_sdM:appendix_model_fits}).

If we average the gravities for the metallicities derived from the model fit independently from
the metal class (i.e.\ sdM, esdM, usdM), we observe that targets with metallicities of $-$0.5 dex
have on average lower gravities ($\sim$4.9 dex) than more metal-poor objects where mean gravities
lie between 5.2 and 5.3 dex. The difference in gravity between field M dwarfs and M subdwarfs
is below our error bar of 0.5 dex, while the difference between metallicities is five times smaller
than our error bars.

%
%%%%%%%%%%%%%%%%%%%%%%%%%%%%%%%%%%%%%%
%%%%% Figure: SpT vs log(g) %%%%%
%%%%%%%%%%%%%%%%%%%%%%%%%%%%%%%%%%%%%%
%
%

%
%%%%%%%%%%%%%%%%%%%%%%%%%%%%%%%%%%%%%%
%%%%% Figure: SpT vs FeH %%%%%
%%%%%%%%%%%%%%%%%%%%%%%%%%%%%%%%%%%%%%
%
\begin{figure}
  \centering
  \includegraphics[width=\linewidth, angle=0]{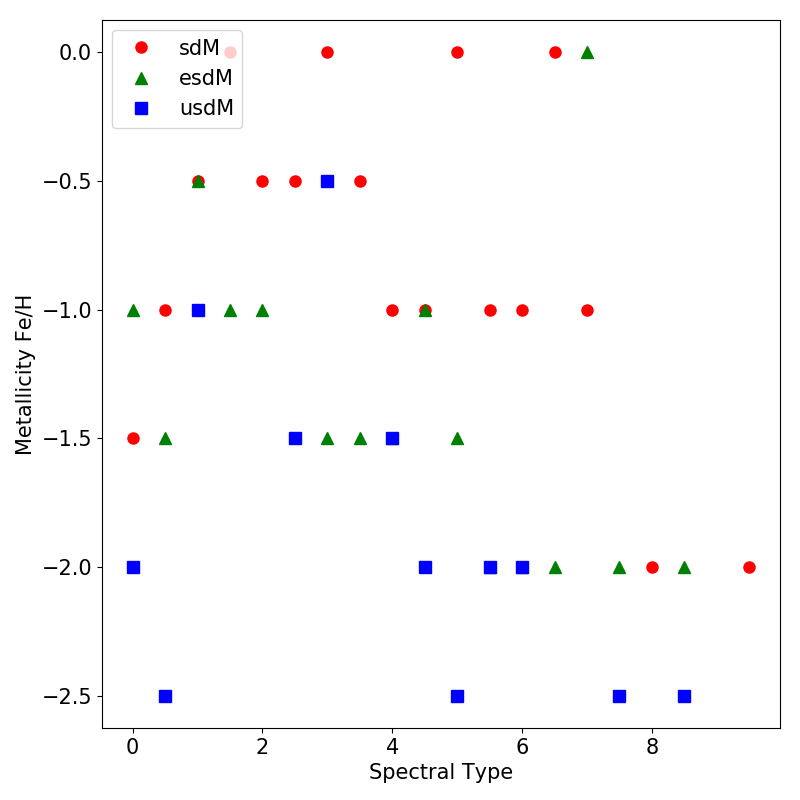}
   \caption{Metallicity as a function of spectral type for sdM (red dots), esdM (green triangles)
   and usdM (blue squares). Error bars on spectral type and metallicity are $\pm$0.5
   and $\pm$0.5 dex, respectively.
}
   \label{fig_MplusTeff_sdM:plot_SpT_FeH}
\end{figure}
\subsection{Metallicity scale}
\label{MplusTeff_sdM:Comparison_MH}

Our work is extending the current metallicity scale of M dwarfs to metallicities below $-$0.5 dex
with an accuracy on the calibration of the order of 0.25--0.5 dex. The classification is available
in the optical for subdwarfs \citep{gizis97a,lepine07c,jao08,savcheva14,kesseli17a} but this is the 
first time this is attempted for a large sample of M0--M9 dwarfs \citep[see also][]{rajpurohit16a}.
The determination of the metallicity scale of subdwarfs is key for several reasons. First, 
the range of metallicities for M subdwarfs is poorly constrained up to now for these objects
that represent the first generation of stars and are key sources to study the chemical evolution
of our Galaxy. On the other hand, M dwarfs are becoming popular to search for low-mass planets 
by the radial velocity and transit techniques, leading independent groups to look at their metallicity.
Nonetheless, all the studies referenced in the introduction focussed on M dwarfs with
metallicity slightly below solar with calibrations accurate to $<$0.15 dex either photometrically
\citep{bonfils05a,johnson09a,schlaufman10a,neves12,hejazi15,dittmann16} or spectroscopically
\citep{woolf05,woolf06,bean06a,woolf09,rojas_ayala10,rojas_ayala12,muirhead12,terrien12a,onehag12,neves13,neves14,hejazi15,newton15a,lindgren16a}.

The number of M dwarf hosts below 0.5 M$_{\odot}$ and metallicities lower than 
$-$0.5 dex is extremely small ($\leq$0.2\%) compared to the 3396 confirmed 
planets\footnote{last update on 10 October 2016 at exoplanetarchive.ipac.caltech.edu}. 
Only three metal-poor M dwarfs have one or several planets: 
Kapteyn \citep[sdM1.0; Fe/H\,=\,$-$0.86 dex;][]{anglada_escude14,robertson15a}, 
GJ\,667 \citep[K3V$+$K5V; Fe/H\,=\,$-$0.55 dex;][]{feroz14,anglada_escude13a,delfosse13}, 
and Kepler\,1124 \citep[Fe/H\,=\,$-$0.59 dex;][]{morton16a}. Another four Kepler
M stars with masses in the range 0.5--0.6 M$_{\odot}$ and metallicities between $-$0.5 and
$-$1.0 harbour planets. 

We present VLT/X-shooter spectra at a resolution of 5400 and 3900 in the VIS and NIR arms, respectively.
We estimate metallicity that rely on models and synthetic spectra whose steps are 0.5 dex,
which we set as our error bars.
Overall, we find that the overall SEDs of subdwarfs, extreme subdwarfs, and ultra-subdwarfs
are best reproduced by metallicities between solar and $-$1.0 dex, $-$1.0 and $-$2.0 dex, and
$-$1.0 and $-$2.5 dex, respectively (Fig.\ \ref{fig_MplusTeff_sdM:plot_SpT_FeH}).
We note that a few sdM have solar metallicities, including the sdM6.5 (1610$-$0040), a
known astrometric binary \citep{koren16} with a mix of spectral features typical of both 
L subdwarfs and solar-type M dwarfs \citep{lepine03b,reiners06a,cushing06}.
We also emphasise that the two coolest subdwarfs (sdM8 and sdM9.5) in our sample have on average 
lower metallicities than subdwarfs earlier than sdM7, suggesting saturation of the metallicity
index \citep{lepine07c} used in the classification of subdwarf \citep{zhang17a}.
We conclude that there is a trend towards lower mean metallicity from subdwarfs ($-$0.5$\pm$0.5 dex)
to extreme subdwarfs ($-$1.5$\pm$0.5 dex), and ultra-subdwarfs ($-$2.0$\pm$0.5 dex), 
consistent with the original definition of these three metal classes.
However, we observe variations from one object to the other, preventing assigning a given M/H
for each metal class. 
We note that the metallicity scale of the line fitting option differs significantly from the
fit of the SEDs, producing lower metallicities by at least 1.0 dex.
(Tables \ref{tab_MplusTeff_sdM:table_best_fits_LL} and \ref{tab_MplusTeff_sdM:table_best_fits_LF}).
We find the same trend if we consider only the optical range (600--1000 nm) for our fitting procedure.

We compared our global results to the recent analysis of \citet{rajpurohit16a} also based on VLT/X-shooter
spectra but for a much smaller sample of subdwarfs: six sdM (five sdM0.5--M3 and one sdM7), two esdM
(esdM2 and esdM4), and a usdM4.5\@. In their Table 2, they list
two possible fits, which usually differ by either 100\,K or 0.5 dex in gravity or metallicity.
Our physical parameters agree with their results within 1--2$\sigma$ of the steps of the 
synthetic SEDs (Table \ref{tab_MplusTeff_sdM:table_param}). The difference can be 
explained by the uncertainties on the spectral types,
typically 0.5 sub-type. We conclude that both T$_{\rm eff}$ vs.\ spectral type relations are
consistent within current error bars on observational and theoretical sides. Moreover,
as stated in  \citet{rajpurohit16a}, these results are fully consistent with the physical
parameters derived from higher resolution optical spectroscopy \citep{rajpurohit14}.

We can also compare our work to the physical parameters of the metal-poor binary 
G\,224-58\,AB composed of a esdK5 and a esdM5.5\@. \citet{pavlenko15} derived the
temperature and gravity for the secondary (T$_{\rm eff}$\,=\,3200$\pm$100 K, 
$\log$g\,=\,5.0$\pm$0.5), and inferred the metallicity from the primary ([Fe/H]\,=\,$-$1.92 dex). 
Their parameters are in good agreement with our values for the esdM5.5 (SDSS\,J09030795$+$0842432) 
in our sample: T$_{\rm eff}$\,=\,3300$\pm$100\,K, $\log$g\,=\,5.5$\pm$0.5 dex, and 
[Fe/H]\,=\,$-$1.5$\pm$0.5 dex (Table \ref{tab_MplusTeff_sdM:table_param}).
Their abundance analysis suggests that many elements (e.g.\ calcium; [Ca/H]\,=\,$-$1.39$\pm$0.03)
are over-abundant compared to iron. We do also find that, although the BT-Settl SEDs reproduce
well the molecular bands of the observed spectra, they fail to fit the atomic lines of many elements 
like Fe, K, Na, and Ca. In other words, the atomic lines from the BT-Settl models are too broad and 
over-estimate the abundances of single elements (see Sect.\ \ref{MplusTeff_sdM:Comparison_Abund}). 
The disappearance of these elements in other molecules (e.g.\ CaOH) could also explain the observed 
discrepancy.

\subsection{Masses}
\label{MplusTeff_sdM:Comparison_Mass}

We inferred model-dependent masses for our sample of subdwarfs from their temperatures looking
at the predictions of the NextGen models \citep{baraffe97}, assuming ages of 10 Gyr for these
old objects. Although this approach is
not correct due to the different inputs physics in the NextGen and BT-Settl grids
\citep[computation available on France Allard's webpage;][]{baraffe15}, it provides estimates on
the masses of these low-metallicity M dwarfs. We derived masses below 0.35 M$_{\odot}$ for 
the most massive M subdwarfs down to $\sim$0.087 M$_{\odot}$ for the coolest objects in our sample, 
i.e.\ close to the stellar/substellar boundary with uncertainties of the order of 10\%
(Table \ref{tab_MplusTeff_sdM:table_param}). 
We note that the gravities quoted by the \citet{baraffe97} models are in the range
$\log$(g)\,=\,5.0--5.5 dex, consistent with the aforementioned gravities derived from 
the synthetic SEDs within the error bars of 0.5 dex.

%
%%%%%%%%%%%%%%%%%%%%%%%%%%%%%%%%%%%%%%%%%%%%%%%%%%%%%%%%%%%%%%
%%%%% Table: Elements for Abundances %%%%%
%%%%%%%%%%%%%%%%%%%%%%%%%%%%%%%%%%%%%%%%%%%%%%%%%%%%%%%%%%%%%%
%
\begin{table}
\centering
\caption[]{Spectral regions, elements, and their numbers of lines considered to derive 
abundances of our sample of subdwarfs}
\begin{tabular}{@{\hspace{1mm}}c c c@{\hspace{1mm}}}
 \hline
 \hline
Wavelength range   & Elements  & \# lines \\
 \hline
5750--5995\AA{}    & Na, Ti, Ca, Fe &  2, 1, 1, 1 \\
7640--7730\AA{}    & K              &  2 \\
8080--8350\AA{}    & Na, Fe         &  3, 1 \\
8300--8850\AA{}    & Ca, Ti, Fe     &  3, 10, 7 \\
 \hline
 \label{MplusTeff_sdM:elements_for_abundance}
\end{tabular}
\end{table}
%

%
%%%%%%%%%%%%%%%%%%%%%%%%%%%%%%%%%%%%%%%%%%%%%%%%%%%%
%%%%% Table: Abundances %%%%%
%%%%%%%%%%%%%%%%%%%%%%%%%%%%%%%%%%%%%%%%%%%%%%%%%%%%
%
\begin{table}
\caption[]{Abundances for five elements (sodium, potassium, calcium, titanium, and iron)
for each subtype and each metal class for a model atmosphere with a fixed temperature structure.
Uncertainties are rounded to the nearest integer and given for a change of $\pm$100\,K 
in effective temperature.
Targets marked with an asterisk indicate low signal-to-noise spectra, whose results must 
be treated with caution.
}
\begin{tabular}{@{\hspace{0mm}}l @{\hspace{2mm}}c @{\hspace{2mm}}c @{\hspace{2mm}}c @{\hspace{2mm}}c @{\hspace{2mm}}c @{\hspace{2mm}}c@{\hspace{0mm}}}
\hline
\hline
SpT           &   Model       &   Na   &   K  &   Ca   &  Ti   &  Fe  \cr
\hline
sdM0.5        & 3700-4.5-1.0 & 0.3$_{-0.2}^{+0.3}$ & 0.7$_{-0.3}^{+0.4}$ & 1.2$_{-0.0}^{+0.0}$ & 0.9$_{-0.1}^{+0.1}$ & 0.9$_{-0.0}^{+0.1}$ \cr
sdM1.0        & 3700-5.0-0.5 & 0.2$_{-0.2}^{+0.2}$ & 0.3$_{-0.3}^{+0.3}$ & 0.5$_{-0.2}^{+0.1}$ & 0.4$_{-0.1}^{+0.1}$ & 0.4$_{-0.1}^{+0.1}$ \cr
sdM1.5$^{*}$  & 3400-4.5-1.5 & 0.4$_{-0.1}^{+0.1}$ & 0.6$_{-0.2}^{+0.2}$ & 1.0$_{-0.0}^{+0.0}$ & 0.5$_{-0.1}^{+0.1}$ & 0.5$_{-0.1}^{+0.1}$ \cr
sdM2.0$^{*}$  & 3500-4.5-1.0 & 0.7$_{-0.2}^{+0.2}$ & 1.0$_{-0.3}^{+0.3}$ & 0.9$_{-0.0}^{+0.0}$ & 0.5$_{-0.1}^{+0.1}$ & 0.5$_{-0.1}^{+0.1}$ \cr
sdM2.5$^{*}$  & 3500-4.5-1.0 & 0.6$_{-0.2}^{+0.2}$ & 0.9$_{-0.3}^{+0.3}$ & 0.7$_{-0.0}^{+0.0}$ & 0.6$_{-0.1}^{+0.1}$ & 0.5$_{-0.1}^{+0.1}$ \cr
sdM3.0        & 3500-5.0-0.0 & $-$0.2$_{-0.1}^{+0.1}$ & 0.6$_{-0.2}^{+0.2}$ & 0.2$_{-0.0}^{+0.0}$ & 0.0$_{-0.1}^{+0.1}$ & 0.1$_{-0.0}^{+0.0}$ \cr
sdM3.5        & 3400-5.0-0.5 & 0.2$_{-0.2}^{+0.2}$ & 0.3$_{-0.3}^{+0.2}$ & 0.3$_{-0.0}^{+0.1}$ & 0.3$_{-0.2}^{+0.2}$ & 0.1$_{-0.1}^{+0.1}$ \cr
sdM4.0        & 3400-5.5-0.5 & $-$0.1$_{-0.2}^{+0.3}$ & 0.2$_{-0.3}^{+0.3}$ & 0.8$_{-0.0}^{+0.0}$ & 0.4$_{-0.1}^{+0.2}$ & 0.2$_{-0.1}^{+0.1}$ \cr
sdM4.5        & 3100-4.5-1.5 & 0.5$_{-0.2}^{+0.2}$ & 0.6$_{-0.3}^{+0.3}$ & 0.6$_{-0.0}^{+0.0}$ & 0.9$_{-0.3}^{+0.3}$ & 0.4$_{-0.1}^{+0.1}$ \cr
sdM5.0$^{*}$  & 3200-5.5-0.0 & 0.5$_{-0.2}^{+0.2}$ & 0.3$_{-0.3}^{+0.4}$ & 0.0$_{-0.0}^{+0.0}$ & 0.3$_{-0.2}^{+0.3}$ & 0.1$_{-0.1}^{+0.2}$ \cr
sdM5.5        & 3200-5.5-1.0 & 0.0$_{-0.2}^{+0.2}$ & 0.1$_{-0.3}^{+0.3}$ & 0.4$_{-0.0}^{+0.2}$ & 0.6$_{-0.2}^{+0.2}$ & 0.1$_{-0.1}^{+0.1}$ \cr
sdM6.0        & 3000-4.5-1.5 & 0.6$_{-0.2}^{+0.2}$ & 0.3$_{-0.3}^{+0.4}$ & 1.4$_{-0.0}^{+0.0}$ & 0.7$_{-0.2}^{+0.3}$ & 0.4$_{-0.1}^{+0.2}$ \cr
sdM6.5$^{*}$  & 2900-5.5-0.0 & 0.2$_{-0.4}^{+0.5}$ & 0.4$_{-0.5}^{+0.6}$ & 0.0$_{-0.0}^{+0.0}$ & 0.0$_{-0.2}^{+0.3}$ & 0.0$_{-0.1}^{+0.2}$ \cr
sdM8.0        & 2800-5.0-2.5 & 0.4$_{-0.2}^{+0.2}$ & 0.3$_{-0.2}^{+0.2}$ & 0.7$_{-0.0}^{+0.0}$ & 0.6$_{-0.2}^{+0.2}$ & 0.0$_{-0.1}^{+0.1}$ \cr
sdM9.5        & 2600-4.5-2.5 & 0.6$_{-0.2}^{+0.2}$ & 0.4$_{-0.2}^{+0.2}$ & 0.5$_{-0.0}^{+0.0}$ & 0.8$_{-0.2}^{+0.2}$ & 0.5$_{-0.1}^{+0.1}$ \cr
 \hline
esdM0.0 & 3600-4.5-1.5 & 0.2$_{-0.2}^{+0.2}$ & 0.4$_{-0.2}^{+0.3}$ & 0.6$_{-0.0}^{+0.0}$ & 0.6$_{-0.1}^{+0.2}$ & 0.3$_{-0.0}^{+0.1}$ \cr
esdM0.5 & 3600-4.5-1.5 & 0.2$_{-0.2}^{+0.2}$ & 0.6$_{-0.2}^{+0.3}$ & 0.5$_{-0.0}^{+0.0}$ & 0.6$_{-0.1}^{+0.2}$ & 0.3$_{-0.0}^{+0.1}$ \cr
esdM1.0 & 3600-4.5-1.5 & 0.3$_{-0.2}^{+0.2}$ & 0.6$_{-0.2}^{+0.3}$ & 0.6$_{-0.0}^{+0.0}$ & 0.7$_{-0.1}^{+0.2}$ & 0.1$_{-0.0}^{+0.1}$ \cr
esdM1.5 & 3400-4.5-1.5 & 0.2$_{-0.2}^{+0.2}$ & 0.4$_{-0.2}^{+0.3}$ & 0.7$_{-0.0}^{+0.0}$ & 0.7$_{-0.1}^{+0.2}$ & 0.4$_{-0.0}^{+0.1}$ \cr
esdM2.0 & 3400-4.5-1.5 & 0.3$_{-0.2}^{+0.2}$ & 0.5$_{-0.2}^{+0.3}$ & 0.8$_{-0.0}^{+0.0}$ & 0.9$_{-0.1}^{+0.2}$ & 0.5$_{-0.0}^{+0.1}$ \cr
esdM3.0 & 3400-5.0-1.5 & 0.2$_{-0.2}^{+0.2}$ & 0.5$_{-0.2}^{+0.2}$ & 0.7$_{-0.0}^{+0.0}$ & 0.9$_{-0.2}^{+0.2}$ & 0.4$_{-0.1}^{+0.1}$ \cr
esdM3.5 & 3400-5.0-1.5 & 0.0$_{-0.2}^{+0.2}$ & 0.4$_{-0.2}^{+0.2}$ & 0.4$_{-0.0}^{+0.0}$ & 0.7$_{-0.2}^{+0.2}$ & 0.2$_{-0.1}^{+0.1}$ \cr
esdM4.0 & 3200-4.5-2.0 & 0.5$_{-0.2}^{+0.2}$ & 0.7$_{-0.2}^{+0.2}$ & 0.4$_{-0.0}^{+0.0}$ & 0.9$_{-0.2}^{+0.2}$ & 0.4$_{-0.1}^{+0.1}$ \cr
esdM4.5 & 3200-4.5-1.5 & 0.6$_{-0.2}^{+0.2}$ & 0.4$_{-0.2}^{+0.2}$ & 0.9$_{-0.0}^{+0.0}$ & 0.8$_{-0.2}^{+0.2}$ & 0.4$_{-0.1}^{+0.1}$ \cr
esdM5.0 & 3200-4.5-1.5 & 0.4$_{-0.2}^{+0.2}$ & 0.7$_{-0.2}^{+0.3}$ & 0.7$_{-0.0}^{+0.0}$ & 0.9$_{-0.2}^{+0.2}$ & 0.5$_{-0.1}^{+0.1}$ \cr
esdM5.5 & 3100-4.5-2.0 & 0.6$_{-0.2}^{+0.2}$ & 0.8$_{-0.2}^{+0.2}$ & 0.8$_{-0.0}^{+0.0}$ & 1.2$_{-0.3}^{+0.3}$ & 0.7$_{-0.1}^{+0.2}$ \cr
esdM6.0 & 3200-5.0-2.0 & 0.1$_{-0.2}^{+0.2}$ & 0.6$_{-0.2}^{+0.2}$ & 0.4$_{-0.0}^{+0.0}$ & 0.8$_{-0.2}^{+0.2}$ & 0.4$_{-0.1}^{+0.1}$ \cr
esdM6.5 & 3100-4.5-2.0 & 0.3$_{-0.3}^{+0.3}$ & 0.7$_{-0.2}^{+0.2}$ & 0.7$_{-0.0}^{+0.0}$ & 1.0$_{-0.3}^{+0.3}$ & 0.4$_{-0.1}^{+0.1}$ \cr
esdM7.5 & 3000-5.0-2.0 & 0.2$_{-0.2}^{+0.2}$ & 0.6$_{-0.2}^{+0.2}$ & 0.7$_{-0.0}^{+0.0}$ & 0.8$_{-0.2}^{+0.2}$ & 0.3$_{-0.1}^{+0.1}$ \cr
esdM8.5 & 2800-4.5-2.5 & 0.4$_{-0.2}^{+0.2}$ & 0.6$_{-0.2}^{+0.2}$ & 0.6$_{-0.0}^{+0.0}$ & 0.8$_{-0.2}^{+0.3}$ & 0.1$_{-0.1}^{+0.1}$ \cr
 \hline
usdM0.0       & 3500-4.5-2.5 & 0.5$_{-0.3}^{+0.4}$ & 1.2$_{-0.4}^{+0.5}$ & 1.2$_{-0.0}^{+0.0}$ & 0.6$_{-0.2}^{+0.3}$ & 0.2$_{-0.2}^{+0.2}$ \cr
usdM0.5$^{*}$ & 3600-5.0-2.5 & $-$0.4$_{-0.2}^{+0.3}$ & 0.6$_{-0.4}^{+0.4}$ & 0.4$_{-0.0}^{+0.0}$ & 0.5$_{-0.2}^{+0.2}$ & 0.1$_{-0.1}^{+0.1}$ \cr
usdM1.0       & 3600-5.0-1.5 & $-$0.2$_{-0.2}^{+0.3}$ & 0.6$_{-0.4}^{+0.4}$ & 0.5$_{-0.0}^{+0.0}$ & 0.3$_{-0.2}^{+0.2}$ & 0.2$_{-0.1}^{+0.1}$ \cr
usdM2.5       & 3400-4.5-2.0 & 0.1$_{-0.2}^{+0.2}$ & 0.8$_{-0.4}^{+0.4}$ & 1.0$_{-0.0}^{+0.0}$ & 0.7$_{-0.2}^{+0.2}$ & 0.8$_{-0.1}^{+0.1}$ \cr
usdM3.0       & 3400-4.5-2.0 & $-$0.2$_{-0.2}^{+0.2}$ & 1.2$_{-0.4}^{+0.4}$ & 0.9$_{-0.0}^{+0.0}$ & 0.5$_{-0.2}^{+0.4}$ & 0.3$_{-0.1}^{+0.1}$ \cr
usdM4.0$^{*}$ & 3500-5.5-2.0 & 0.1$_{-0.2}^{+0.3}$ & 0.8$_{-0.3}^{+0.4}$ & 0.9$_{-0.0}^{+0.0}$ & 0.8$_{-0.2}^{+0.2}$ & 0.5$_{-0.1}^{+0.1}$ \cr
usdM4.5       & 3300-5.0-2.0 & $-$0.1$_{-0.2}^{+0.3}$ & 0.6$_{-0.3}^{+0.3}$ & 0.5$_{-0.0}^{+0.0}$ & 0.5$_{-0.2}^{+0.2}$ & 0.3$_{-0.1}^{+0.1}$ \cr
usdM5.0       & 3400-5.5-2.5 & $-$0.4$_{-0.2}^{+0.2}$ & 0.6$_{-0.3}^{+0.3}$ & 0.6$_{-0.0}^{+0.0}$ & 0.5$_{-0.2}^{+0.2}$ & 0.3$_{-0.1}^{+0.1}$ \cr
usdM6.0       & 3200-5.0-2.0 & 0.1$_{-0.1}^{+0.2}$ & 1.0$_{-0.3}^{+0.3}$ & 0.4$_{-0.0}^{+0.0}$ & 0.6$_{-0.1}^{+0.1}$ & 0.3$_{-0.1}^{+0.1}$ \cr
usdM7.5       & 3000-5.0-2.5 & $-$0.1$_{-0.2}^{+0.2}$ & 1.0$_{-0.3}^{+0.3}$ & 0.4$_{-0.1}^{+0.1}$ & 0.6$_{-0.1}^{+0.1}$ & 0.2$_{-0.1}^{+0.2}$ \cr
usdM8.5       & 3000-5.5-2.5 & $-$0.2$_{-0.2}^{+0.2}$ & 0.8$_{-0.3}^{+0.3}$ & 0.3$_{-0.1}^{+0.1}$ & 0.4$_{-0.2}^{+0.2}$ & 0.0$_{-0.1}^{+0.2}$ \cr
 \hline
 \label{MplusTeff_sdM:results_abundance}
\end{tabular}
\end{table}
%

%
%%%%%%%%%%%%%%%%%%%%%%%%%%%%%%%%%%%%%%
%%%%% Figure %%%%%
%%%%%%%%%%%%%%%%%%%%%%%%%%%%%%%%%%%%%%
%
\begin{figure*}
\centering
  \includegraphics[width=0.48\linewidth, angle=0]{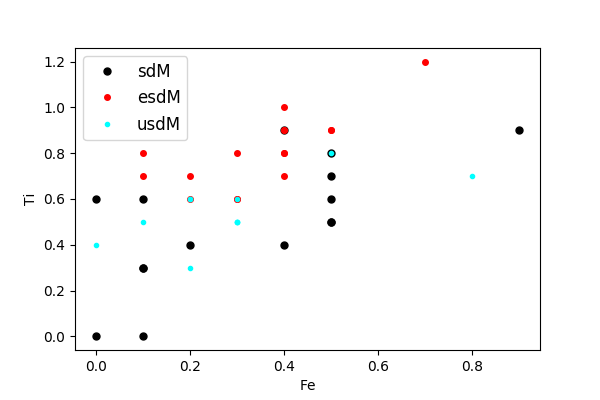}
  \includegraphics[width=0.48\linewidth, angle=0]{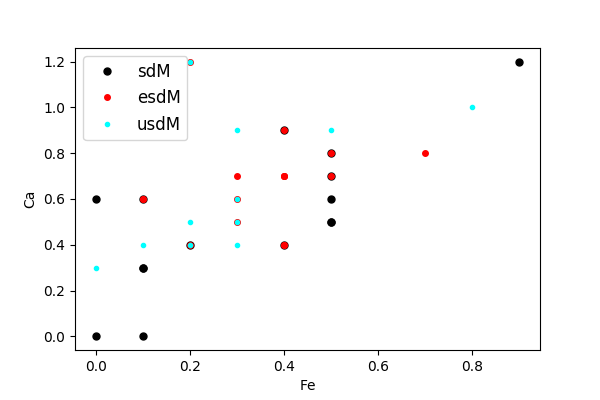}
  \includegraphics[width=0.48\linewidth, angle=0]{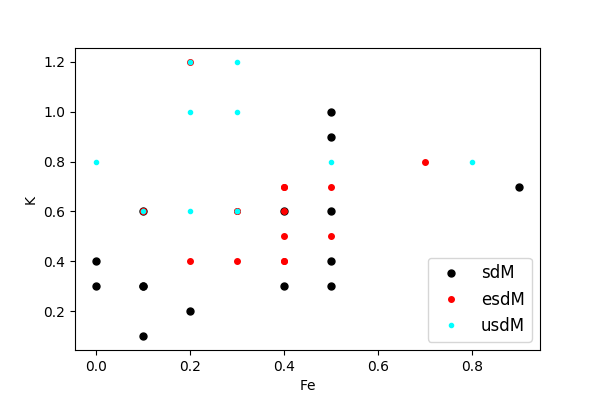}
  \includegraphics[width=0.48\linewidth, angle=0]{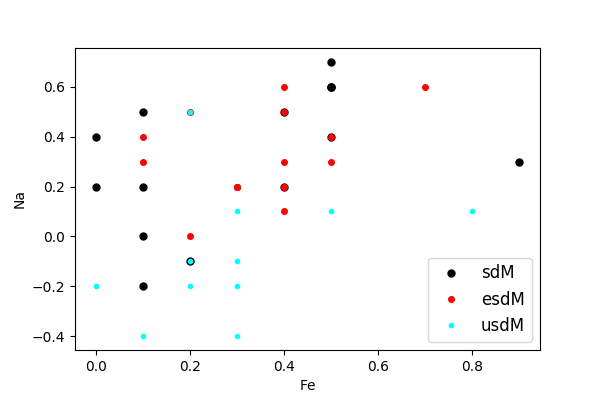}
  \includegraphics[width=0.48\linewidth, angle=0]{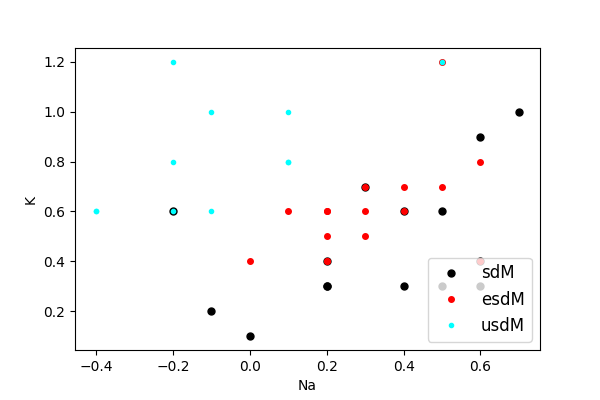}
\caption{Plots showing the scatter in the abundances of subdwarfs (black symbols), extreme 
subdwarfs (red), and ultra-subdwarfs (cyan) for several elements: Fe, Ti, K, Na, and Ca
(Table \ref{MplusTeff_sdM:elements_for_abundance}).
}
\label{fig_MplusTeff_sdM:hists_abundances}
\end{figure*}
\subsection{Abundances}
\label{MplusTeff_sdM:Comparison_Abund}

We decided to work with the temperatures, gravities, and metallicities derived from the best fits
to the optical regions only (Table \ref{tab_MplusTeff_sdM:table_best_fits_VV} in Appendix \ref{MplusTeff_sdM:appendix_model_fits}) to derive the abundances of several elements
(Table \ref{MplusTeff_sdM:elements_for_abundance}). The relatively low resolution and reduced signal-to-noise
of the observed spectra limit the accuracy of our results. Moreover, we work here with metal-deficient stars, 
in which spectra 
are not critically affected by blending effects. We carried out the minimisation procedure using the
PHOENIX model atmospheres to compute synthetical spectra with the WITA program \cite{pavlenko97a}.
In these computations, we accounted for the main molecular opacity sources, as described in
\citet{pavlenko15} and \citet{pavlenko14}. We employed the atomic line information from the
the VALD2 and VALD3 databases \citep{kupka99,ryabchikova11}.
We selected a few spectral ranges with well defined atomic lines and fitted them in framework of 
the LTE approach to determine abundances of several elements listed in
Table \ref{MplusTeff_sdM:elements_for_abundance}.

To determine the abundances, we compared the residuals of the fluxes and reduced to the local continuum 
or pseudo-continuum of the observed spectra. We determined these continua/pseudo-continua
for all spectral ranges given in Table \ref{MplusTeff_sdM:elements_for_abundance}.
We carried out the computations for 20 points abundance grid with a step 0.1 dex around the mean 
abundance for each metal and spectral subclass.
The abundances correspond to over-abundances if they are positive, fixing the iron abundances
to the metallicity derived from the model fit of the optical region, assuming a solar abundance 
of iron of $-$4.55 dex \citep{asplund09}.

When several lines are present in a given spectral range, we averaged the values of each abundance.
We give the uncertainties on the abundances rounded to the nearest integer for a range of
$\pm$100\,K in effective temperature, which represents the error on our model fits (and the step
of the models). We note that the sodium, potassium, and neutral calcium lines are the most sensitive
to small changes in effective temperature due to the low ionisation potentials.
We estimate the formal accuracy of the abundance determination to be $\pm$0.1 dex due to several
uncertainties in the procedure, such as low signal-to-noise ratio and the determination of the
levels of the continuum and pseudo-continuum. We list the final abundances of five elements 
(Na, K, Ca, Ti, and Fe) with their uncertainties in Table \ref{MplusTeff_sdM:results_abundance}. 
Some of the stars, marked
with an asterisk, have low signal-noise spectra so the abundances should be interpreted with caution.

We should note that any significant changes in the abundance of alkali metals affect the fluxes
of the continuum, hence, the strength of the lines. Alkali metals (Na, K, Ca, Mg) are known donors
of free electrons in atmospheres of late-type stars. In our case, we re-computed the continuum
fluxes for all the abundances, so, in a first approach, we accounted for the effect of
opacity changes. However, these changes of abundances may affect the temperature structure of the
model atmosphere, changes not accounted for because we employed model atmospheres with a fixed 
temperature structure.

Our abundance analysis was performed for the model parameters obtained from the fits to 
observed SEDs (Section \ref{MplusTeff_sdM:Models}). Furthermore, resolution of the 
fitted spectra was not high enough to get high quality results. On the second hand, we work here with 
metal-deficient stars, in which spectra are not critically affected by blending effects. We determined 
a few spectral ranges with well defined atomic lines and fitted them in framework of the LTE approach. 
Despite all factors constraining the accuracy of our analysis we get a few confident enough results:

\begin{itemize}
\item [$\bullet$] we see a shift of Ti abundance distribution toward larger metallicities with respect to Fe
for the lowest metallicities (top left panel of Fig.\ \ref{fig_MplusTeff_sdM:hists_abundances}).
\item [$\bullet$] we observe a Ca over-abundance in the atmospheres of all subdwarfs (top right panel of 
Fig.\ \ref{fig_MplusTeff_sdM:hists_abundances}). 
This result agrees with the known enhancement of $\alpha$ elements in the atmospheres of halo dwarfs
\citep[e.g.][]{francois86a,nissen94a,gratton94a,fuhrmann95a}.
\item [$\bullet$] we find that K tends to be over-abundant at higher metallicities, opposite effect to Na
(middle panels of Fig.\ \ref{fig_MplusTeff_sdM:hists_abundances}). This difference increases
with lower metallicity and is most notable for ultra-subdwarfs.
Perhaps, these elements have different chemical histories. The difference is well seen in the comparison 
plot of the distribution of both species where the three metal classes are fairly well separated 
(bottom panel of Fig.\ \ref{fig_MplusTeff_sdM:hists_abundances}.
\item [$\bullet$] we cannot describe the abundances in the atmospheres of our targets with one parameter 
metallicity (Table \ref{MplusTeff_sdM:results_abundance} and Fig.\ \ref{fig_MplusTeff_sdM:hists_abundances}).
\end{itemize}
%

%
%%%%%%%%%%%%%%%%%%%%%%%%%%%%%%%%%%%%%%
%%%%% Discussion %%%%%
%%%%%%%%%%%%%%%%%%%%%%%%%%%%%%%%%%%%%%
%
\section{Discussion}
\label{MplusTeff_sdM:discussion}

We observe that on average the optical range of the subdwarf spectra yields cooler temperatures, lower
gravities, and metallicities. Hence, the availability of the near-infrared slope is quite important to determine
the physical parameters of the three metal classes and break down the temperature/metallicity degeneracy present
in the analysis of the sole visible range. The temperature sequence is well determined for all
metal class, with the earliest spectral types being the warmest cases. The metallicity is on average lower
for the ultra-subdwarfs, with all three metal class being more metal poor than solar as expected.
Those results are, however, global with the physical parameters sensitive to steps of the models in
temperature, gravity, and metallicity. Smaller changes in those parameters are most likely limited
by the quality (signal-to-noise ratio) and resolution of the X-shooter spectra.

We checked the few cases where our best fits suggest solar metallicity. There are four sdM in this case 
(including the sdM6.5 specifically treated below) and only one esdM (esdM7.0) also discussed below. 
For the sdM1.5 we find that the best ``FF'' fits indicate solar metallicity while the other procedures 
suggest $-$0.5 dex. The inspection of the width of the lines (K, Na, Ca) is inconclusive to decide
on one or the other metallicity (Fig.\ \ref{fig_MplusTeff_sdM:full_XSH_model_lines_sdM}). 
For the sdM3.0, all four fits are identical and suggest solar metallicity while the third best fit of 
the ``FF'' procedure yields $-$0.5 dex. The case of sdM5.0 indicates solar metallicity too but all other
fits suggest $-$0.5 dex, confirmed by the width of the sodium and potassium lines, which advocate 
metallicity lower than solar with a temperature 3200\,K\@. In the case of these three subdwarfs, 
we conclude that they might be peculiar somehow and require further investigation with better quality data.

We also reviewed the four extreme subdwarfs (esdM1.0, esdM1.5, esdM2.0, and esdM4.5) with differences of 
200\,K or more in temperature and $\geq$ 1.0 dex in metallicity comparing the four fitting procedure
described in Section \ref{MplusTeff_sdM:Comparison_chi2}. We note that the three best fits of the 
``FF'' procedure yield results within the step of the models. Therefore, we inspected the sodium, 
potassium, and calcium lines where we noticed that the best model fits occurred for the lowest 
metallicity values in all cases (Fig.\ \ref{fig_MplusTeff_esdM:full_XSH_model_lines_esdM}). 

Finally, we observe four notable outliers discussed below after checking the best fits derived from 
the different procedures. We emphasise that we obtained one single spectrum per target.
\begin{itemize}
\item {\it{sdM0.0}} is not included in our abundance analysis because the sodium and potassium lines are 
clearly resolved in our X-shooter spectrum. We confirm this target as a spectroscopic binary, which will 
be discussed in a separate paper. Any fit to spectral lines and abundance analysis on the spectrum
taken at this specific epoch is therefore flawed.
\item {\it{sdM6.5}} (1610$-$0040) is a known astrometric binary \citep{koren16} with a mix of spectral features 
typical of both L subdwarfs and solar-type M dwarfs \citep{lepine03b,reiners06a,cushing06}. We derive a
temperature of 2900\,K with a gravity of 5.5 dex and solar metallicity with a reasonable ``FF'' model fit
(Fig.\ \ref{fig_MplusTeff_sdM:full_XSH_model_FF_sdM}).
\item {\it{esdM7.0}} show solar metallicity in the case of the full fitting procedure but lower metallicities 
with the other three cases. We inspected the sodium, potassium, and calcium lines of this object and conclude 
that the lines are broader than model predictions at solar metallicity, favouring the metal poor solution.
(Fig.\ \ref{fig_MplusTeff_esdM:full_XSH_model_lines_esdM}). We note that this object has a low quality 
spectrum too.
\item {\it{usdM3.0}} exhibits a large difference when inspecting the best three chi$^{2}$ fits using the 
``FF'' procedure, with variations of up to 200\,K in temperature and 2.0 dex in metallicity. 
These differences are larger than any other object in our sample. We inspected the width and depth 
of the main sodium and potassium lines and find that they appear shallower with a potential double peak, 
which may indicate binarity (Fig.\ \ref{fig_MplusTeff_usdM:full_XSH_model_lines_usdM}). However, we
obtained one single spectrum for this source so we cannot exclude other
phenomenon like rotation, flare activity or spots that can affect our results.
A few additional spectra with a minimum spectral resolution of 10000 are required to confirm this possibility,
which would explain the large variations in the determination of the physical parameters.
\end{itemize}
%

%
%%%%%%%%%%%%%%%%%%%%%%%%%%%%%%%%%%%%%%
%%%%% Conclusions %%%%%
%%%%%%%%%%%%%%%%%%%%%%%%%%%%%%%%%%%%%%
%
\section{Conclusions}
\label{MplusTeff_sdM:conclusions}

Our atlas of VLT/X-shooter 0.45--2.5 $\mu$m moderate-resolution spectra represents an 
important database to classify metal-poor subdwarfs and has a legacy value for future 
large-scale surveys \citep[WEAVE, 4MOST, LSST;][]{dalton12,dalton14,deJong14a,ivezic08a}
and space missions to come such as the James Webb Space Telescope \citep{gardner09} and 
Euclid \citep{mellier16a}. We derived radial velocities for all subdwarfs from the shift of 
the strongest optical lines. We inferred physical parameters (metallicities, temperatures, and 
gravities) for 43 metal-poor M dwarf by comparing their spectral energy distributions over the 
450--2500\,nm range to the latest BT-Settl synthetic spectra. The main results of our analysis are:

\begin{itemize}
\item [$\bullet$] the best gravity range for M subdwarfs is $\log$(g)\,=\,5.0--5.5 dex.
\item [$\bullet$] the metallicities inferred from the BT-Settl models for subdwarfs, extreme subdwarfs,
and ultra-subdwarfs are $-$0.5, $-$1.5, and $-$2.0 dex with errors of 0.5 dex, respectively.
\item [$\bullet$] the ranges in T$_{\rm eff}$ for subdwarfs, extreme subdwarfs, and ultra-subdwarfs are
comparable and lie in the intervals 
3700--2600\,K, 3800--2900\,K, and 3700--2900\,K with uncertainties of 100\,K, respectively.
\item [$\bullet$] the Ca and Ti elements show an over-abundance while Na behaves in an opposite
manner when compared to the iron abundance.
\end{itemize}

Improvements in the determination of the physical parameters and abundances of metal-poor low-mass
dwarfs require refined and uniform/complete model atmosphere grids to improve the fits.
More advanced procedures are needed to improve the quality of the fits to observed SEDs
treating atomic lines in a self-consistent approach, with model atmospheres and spectra are
computed for one set of input parameters.
Models should take into account the enhancement of C/O and improve the modelling of single lines
updating abundances of the various elements present in cool atmospheres.
To truly determine the physical parameters of M subdwarfs and test evolutionary models at low metallicity, 
the discovery of metal-poor low-mass transiting eclipsing binaries is key.

%
%%%%%%%%%%%%%%%%%%%%%%%%%%%%%%%
%%%%%  ACKNOWLEDGEMENTS %%%%%
%%%%%%%%%%%%%%%%%%%%%%%%%%%%%%%
%
\begin{acknowledgements}
NL was funded by the Ram\'on y Cajal fellowship number 08-303-01-02, and supported by
the grants numbers AYA2010-19136 and AYA2015-69350-C3-2-P from Spanish Ministry of
Economy and Competitiveness (MINECO). We thank Victor B\'ejar,
Yakiv Pavlenko, and ZengHua Zhang for constructive comments on this work.
The research leading to these results has received funding from the French Programme National 
de Physique Stellaire and the Programme National de Plan- etologie of CNRS (INSU).
FA thanks financial support from the Fundaci\'on Jes\'us Serra for a 2 month stay (Jan--Feb 2015)
as a visiting professor at the Instituto de Astrof\'isica de Canarias (IAC) in Tenerife.

This work is based on observations collected with X-shooter on the VLT at the 
European Southern Observatory, Chile, under programmes 089.C-0140(A), 091.C-0264(A), 
092.D-0600(A), and 093.C-0610(A).

This research has made use of the Simbad and Vizier databases, operated
at the Centre de Donn\'ees Astronomiques de Strasbourg (CDS), and
of NASA's Astrophysics Data System Bibliographic Services (ADS).

%SDSS\\
    Funding for the Sloan Digital Sky Survey IV has been provided by
    the Alfred P. Sloan Foundation, the U.S. Department of Energy Office of
    Science, and the Participating Institutions. SDSS-IV acknowledges
    support and resources from the Center for High-Performance Computing at
    the University of Utah. The SDSS web site is www.sdss.org.
    SDSS-IV is managed by the Astrophysical Research Consortium for the 
    Participating Institutions of the SDSS Collaboration including the 
    Brazilian Participation Group, the Carnegie Institution for Science, 
    Carnegie Mellon University, the Chilean Participation Group, the French Participation Group, 
Harvard-Smithsonian Center for Astrophysics, 
    Instituto de Astrof\'isica de Canarias, The Johns Hopkins University, 
    Kavli Institute for the Physics and Mathematics of the Universe (IPMU) / 
    University of Tokyo, Lawrence Berkeley National Laboratory, 
    Leibniz Institut f\"ur Astrophysik Potsdam (AIP),  
    Max-Planck-Institut f\"ur Astronomie (MPIA Heidelberg), 
    Max-Planck-Institut f\"ur Astrophysik (MPA Garching), 
    Max-Planck-Institut f\"ur Extraterrestrische Physik (MPE), 
    National Astronomical Observatory of China, New Mexico State University, 
    New York University, University of Notre Dame, 
    Observat\'ario Nacional / MCTI, The Ohio State University, 
    Pennsylvania State University, Shanghai Astronomical Observatory, 
    United Kingdom Participation Group,
    Universidad Nacional Aut\'onoma de M\'exico, University of Arizona, 
    University of Colorado Boulder, University of Oxford, University of Portsmouth, 
    University of Utah, University of Virginia, University of Washington, University of Wisconsin, 
    Vanderbilt University, and Yale University.
\end{acknowledgements}
%

%
%%%%%%%%%%%%%%%%%%%%%%%%%%%%%%%%%%%%%%%
%%%%%%%%  Bibliography  %%%%%%%%
%%%%%%%%%%%%%%%%%%%%%%%%%%%%%%%%%%%%%%%
%
%\begin{thebibliography}{}
\bibliographystyle{aa}
\bibliography{../../AA/mnemonic,../../AA/biblio_old}
%\end{thebibliography}
%

%
%%%%%%%%%%%%%%%%%%%%%%%%%%%%%%%%%%%%%%%%%%
%%%%% Appendix: subdwarf VO archive %%%%%
%%%%%%%%%%%%%%%%%%%%%%%%%%%%%%%%%%%%%%%%%%
%
\begin{appendix}

\section{Model fit to VLT/X-shooter spectra}
\label{MplusTeff_sdM:appendix_model_fits}
%

%
%%%%%%%%%%%%%%%%%%%%%%%%%%%%%%%%%%%%%%%%%
%%%%% Figure: XSH spec vs model: FF %%%%%
%%%%%%%%%%%%%%%%%%%%%%%%%%%%%%%%%%%%%%%%%
%
%%%%% sdM %%%%%
%
\begin{figure*}
  \centering
  \includegraphics[width=\linewidth, angle=0]{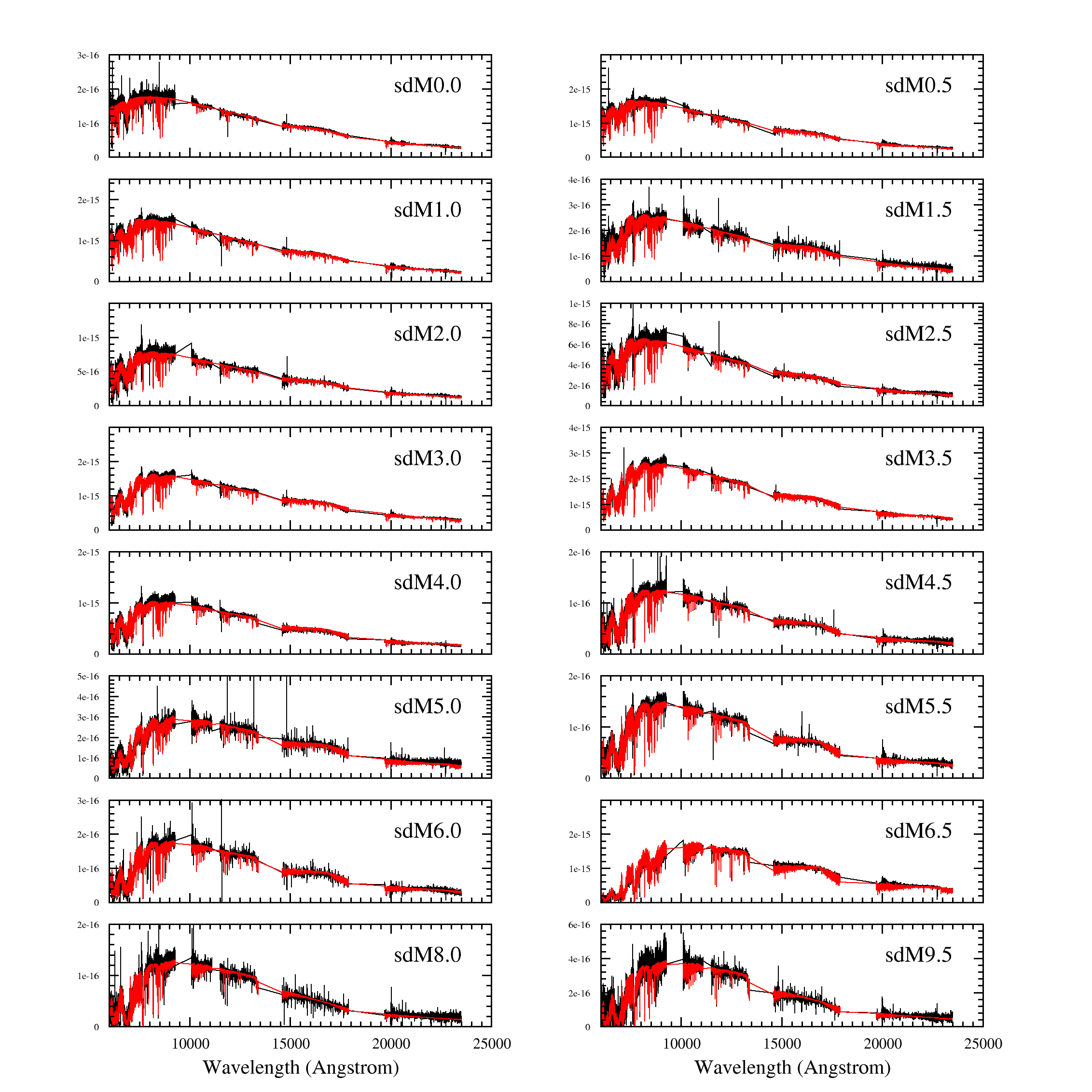}
   \caption{VLT/X-shooter UVB (450--550\,nm), VIS (550--1000 nm) and NIR (1000--2500 nm) spectra 
   of sdM with spectral types from M0.0 to M9.5 (black lines) compared with the best BT-Settl
   spectra smoothed to the observed spectra. Spectral types are quoted in the top right corner
   for each subtype. The fits shown are for the FF case.
   }
   \label{fig_MplusTeff_sdM:full_XSH_model_FF_sdM}
\end{figure*}
%

%
%%%%% esdM %%%%%
%
\begin{figure*}
  \centering
  \includegraphics[width=\linewidth, angle=0]{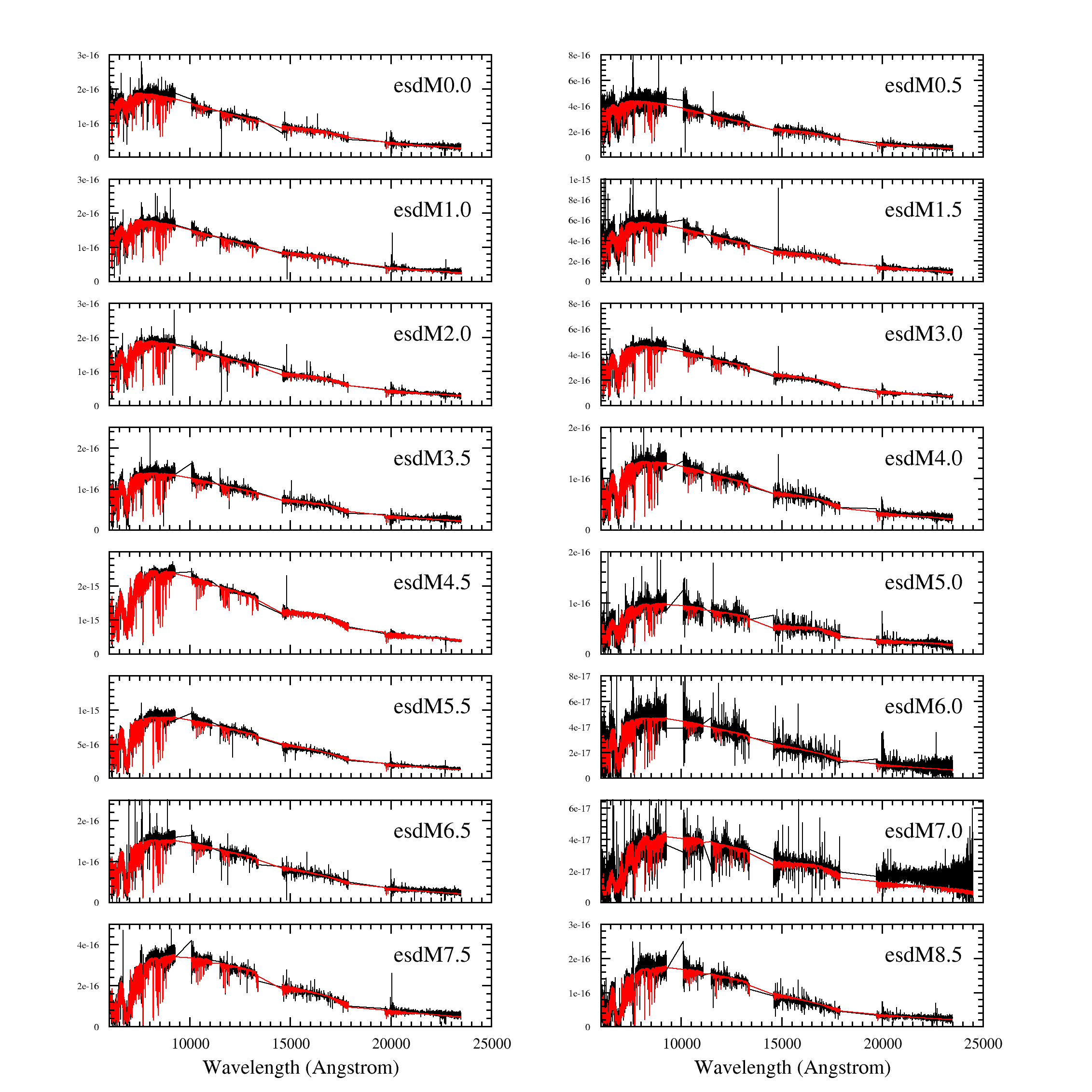}
   \caption{Same as Fig.\ \ref{fig_MplusTeff_sdM:full_XSH_model_FF_sdM} for extreme subdwarfs (esdM).
   }
   \label{fig_MplusTeff_sdM:full_XSH_model_FF_esdM}
\end{figure*}
%

%
%%%%% usdM %%%%%
%
\begin{figure*}
  \centering
  \includegraphics[width=\linewidth, angle=0]{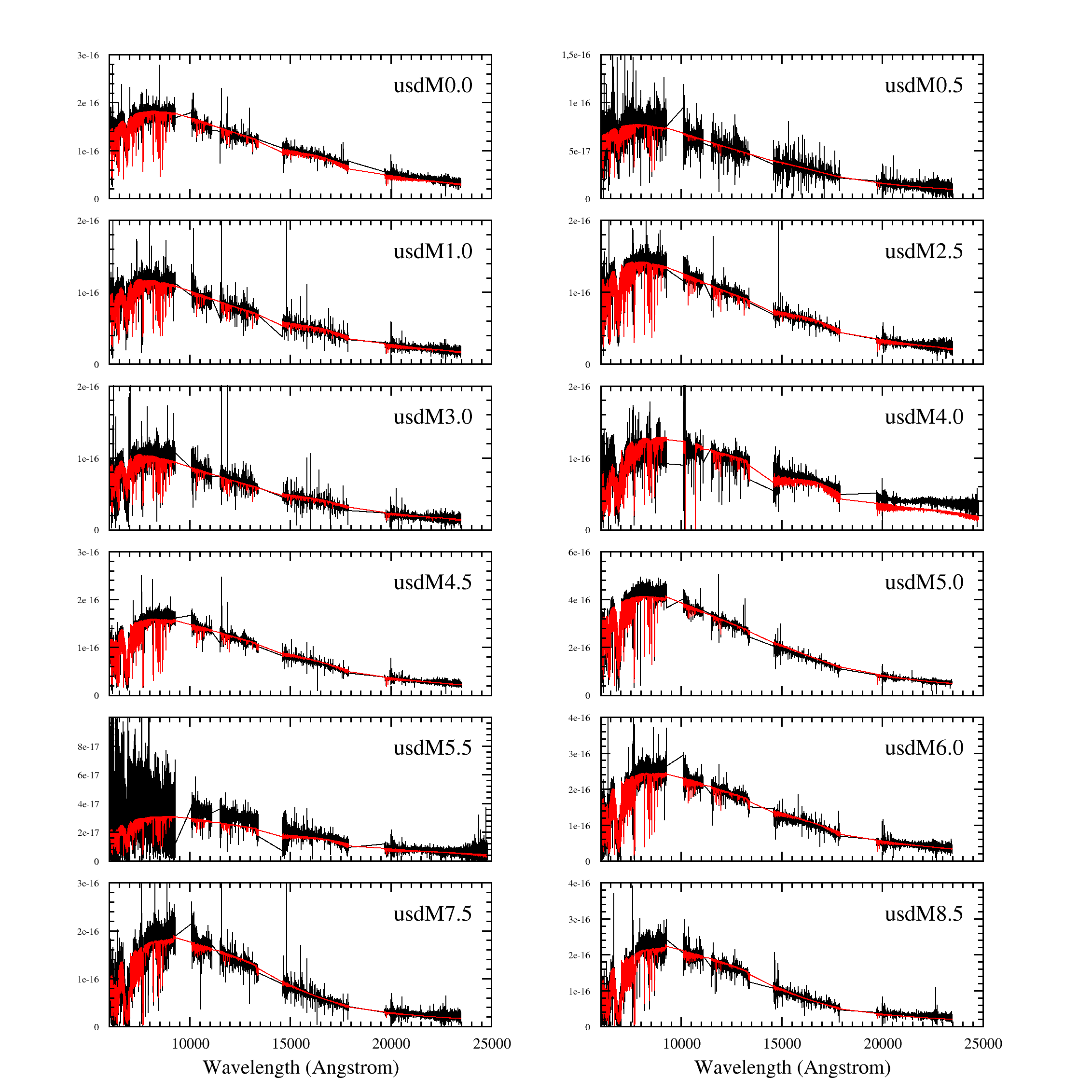}
   \caption{Same as Fig.\ \ref{fig_MplusTeff_sdM:full_XSH_model_FF_sdM} for ultra subdwarfs (usdM).
   }
   \label{fig_MplusTeff_sdM:full_XSH_model_FF_usdM}
\end{figure*}
%

%
%%%%%%%%%%%%%%%%%%%%%%%%%%%%%%%%%%%%%%%%%
%%%%% Figure: XSH spec vs model: LL %%%%%
%%%%%%%%%%%%%%%%%%%%%%%%%%%%%%%%%%%%%%%%%
%
%%%%% sdM %%%%%
%
\begin{figure*}
  \centering
  \includegraphics[width=\linewidth, angle=0]{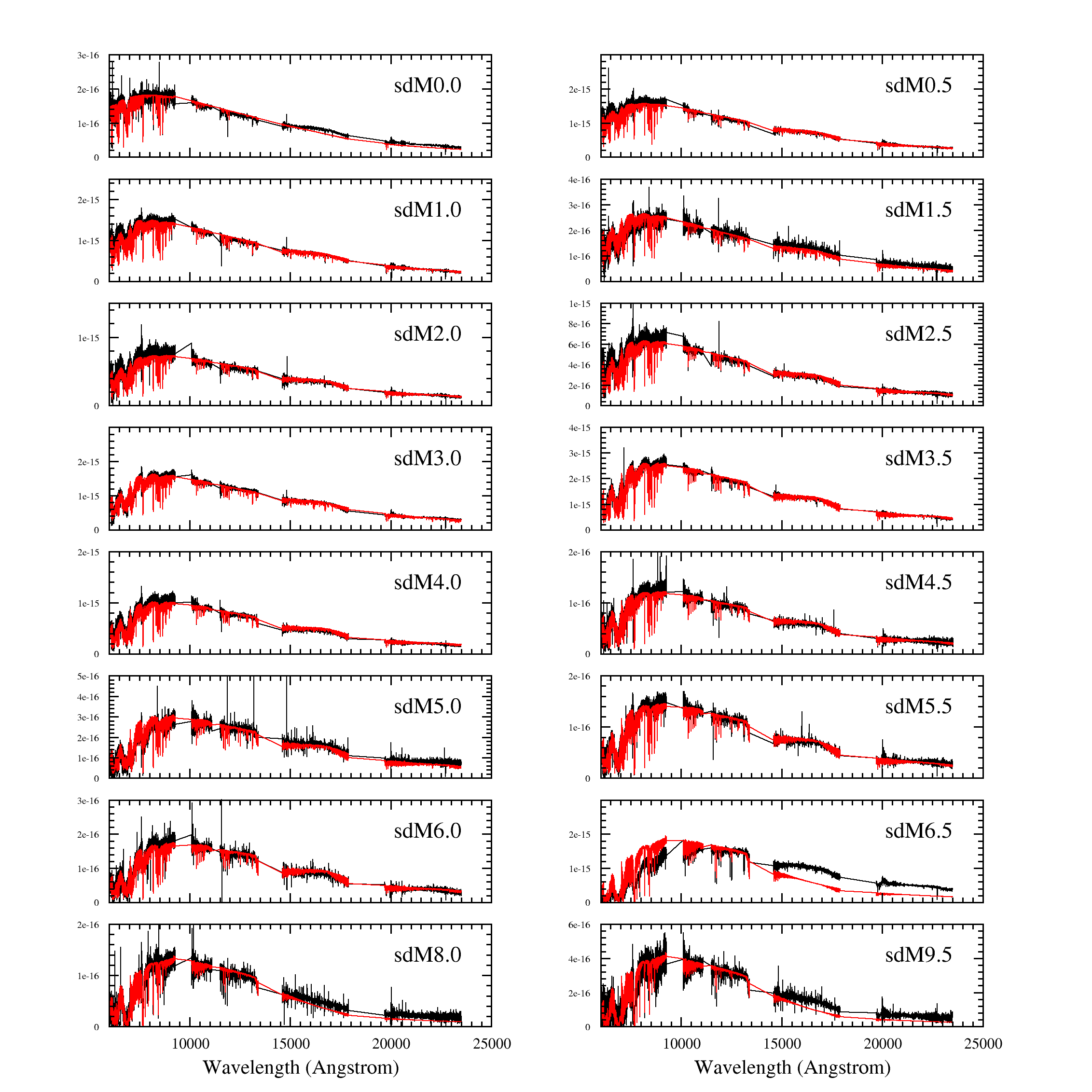}
   \caption{VLT/X-shooter UVB (450--550\,nm), VIS (550--1000 nm) and NIR (1000--2500 nm) spectra
   of sdM with spectral types from M0.0 to M9.5 (black lines) compared with the best BT-Settl
   spectra smoothed to the observed spectra. Spectral types are quoted in the top right corner
   for each subtype. The fits shown are for the LL case.
   }
   \label{fig_MplusTeff_sdM:full_XSH_model_LL_sdM}
\end{figure*}
%

%
%%%%% esdM %%%%%
%
\begin{figure*}
  \centering
  \includegraphics[width=\linewidth, angle=0]{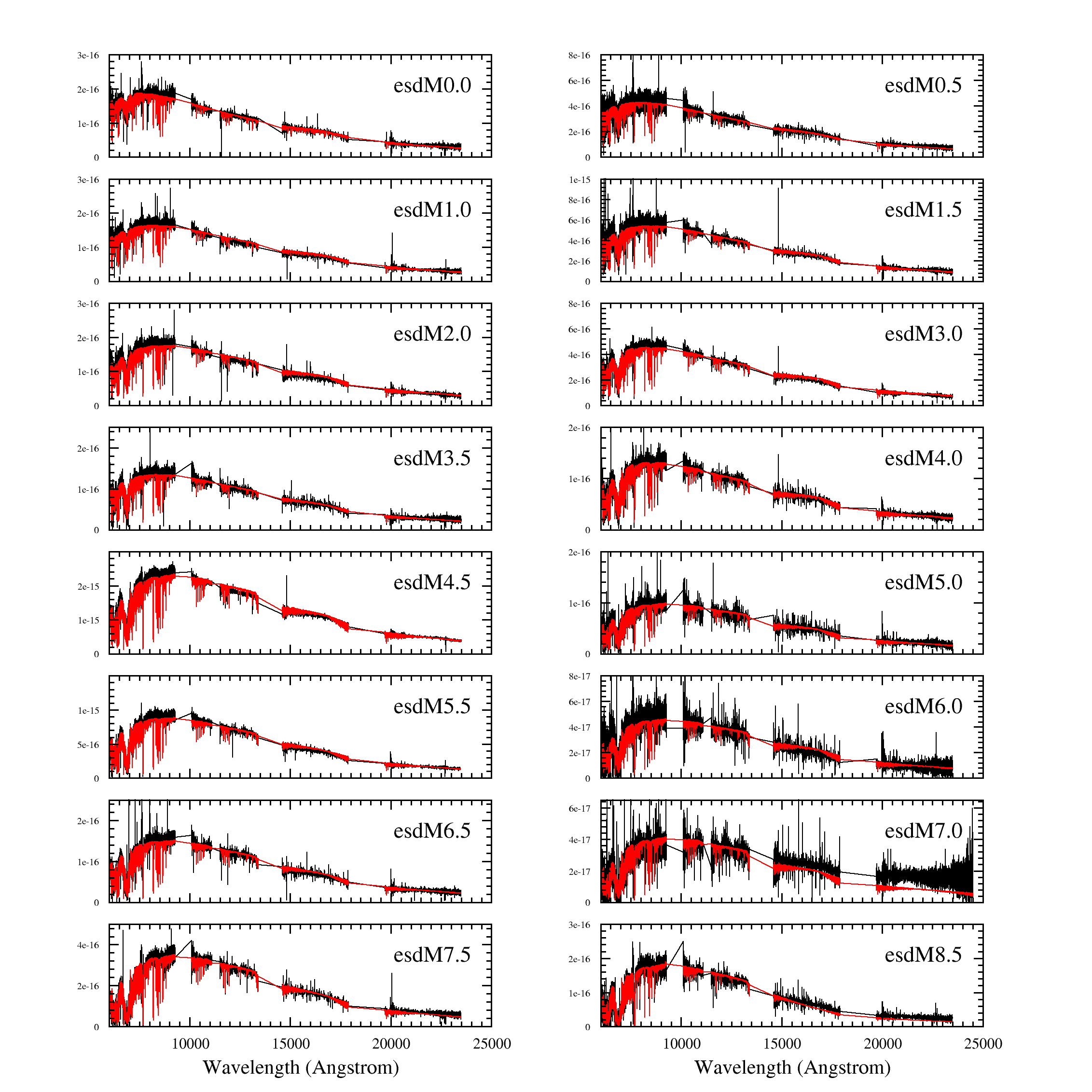}
   \caption{Same as Fig.\ \ref{fig_MplusTeff_sdM:full_XSH_model_LL_sdM} for extreme subdwarfs (esdM).
   }
   \label{fig_MplusTeff_sdM:full_XSH_model_LL_esdM}
\end{figure*}
%

%
%%%%% usdM %%%%%
%
\begin{figure*}
  \centering
  \includegraphics[width=\linewidth, angle=0]{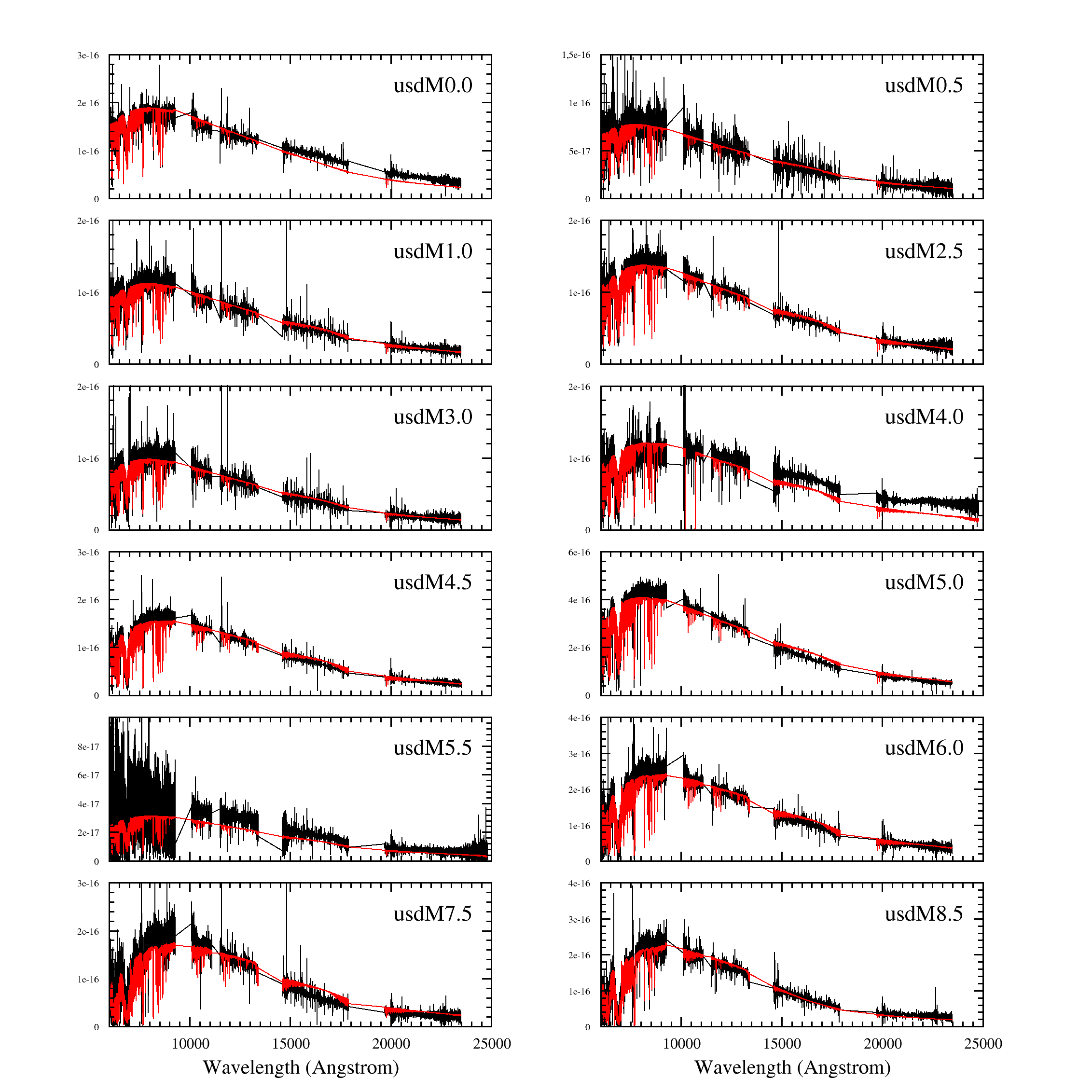}
   \caption{Same as Fig.\ \ref{fig_MplusTeff_sdM:full_XSH_model_LL_sdM} for ultra subdwarfs (usdM).
   }
   \label{fig_MplusTeff_sdM:full_XSH_model_LL_usdM}
\end{figure*}
%

%
%%%%%%%%%%%%%%%%%%%%%%%%%%%%%%%%%%%%%%%%%
%%%%% Figure: XSH spec vs model: FL %%%%%
%%%%%%%%%%%%%%%%%%%%%%%%%%%%%%%%%%%%%%%%%
%
%%%%% sdM %%%%%
%
\begin{figure*}
  \centering
  \includegraphics[width=\linewidth, angle=0]{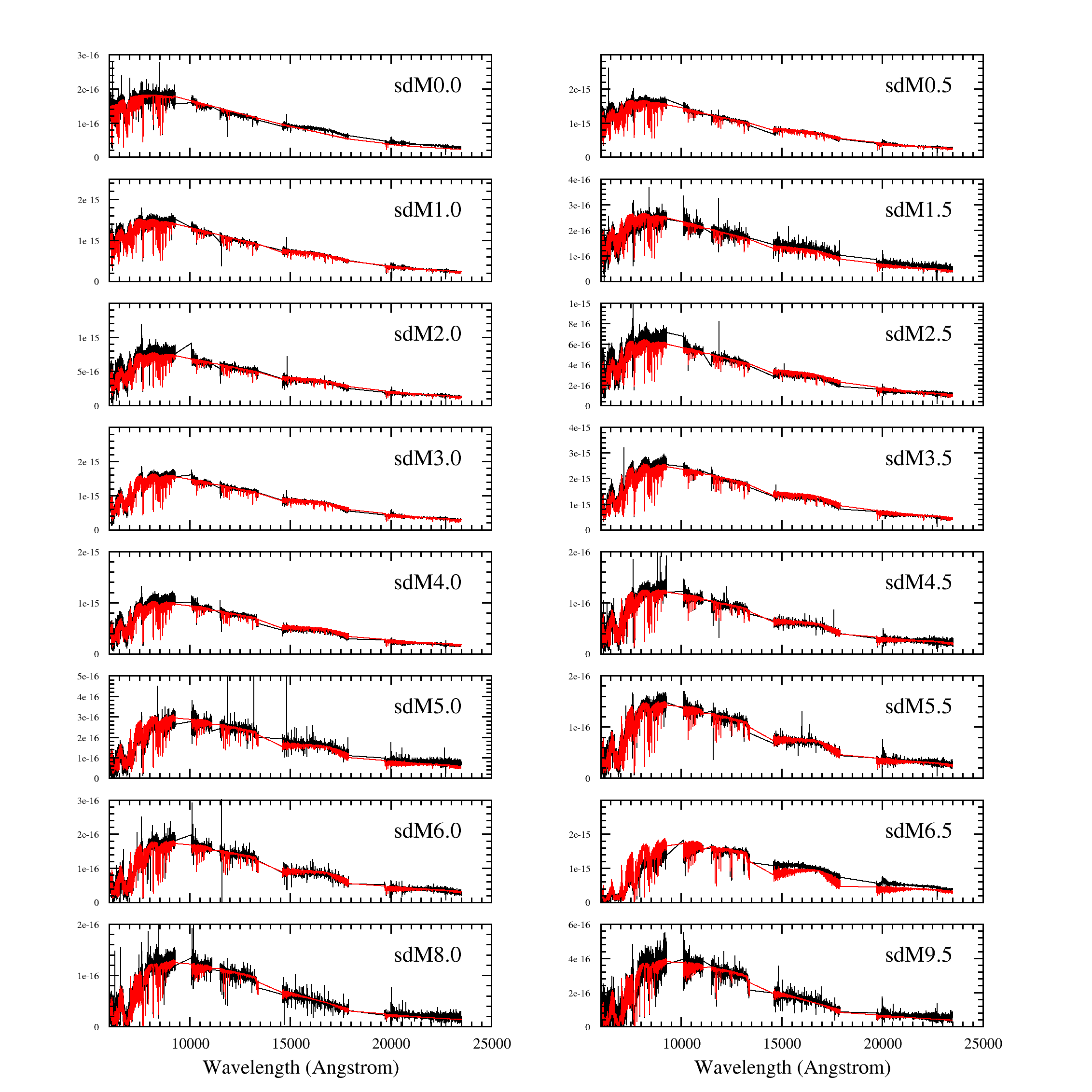}
   \caption{VLT/X-shooter UVB (450--550\,nm), VIS (550--1000 nm) and NIR (1000--2500 nm) spectra
   of sdM with spectral types from M0.0 to M9.5 (black lines) compared with the best BT-Settl
   spectra smoothed to the observed spectra. Spectral types are quoted in the top right corner
   for each subtype. The fits shown are for the FL case.
   }
   \label{fig_MplusTeff_sdM:full_XSH_model_FL_sdM}
\end{figure*}
%

%
%%%%% esdM %%%%%
%
\begin{figure*}
  \centering
  \includegraphics[width=\linewidth, angle=0]{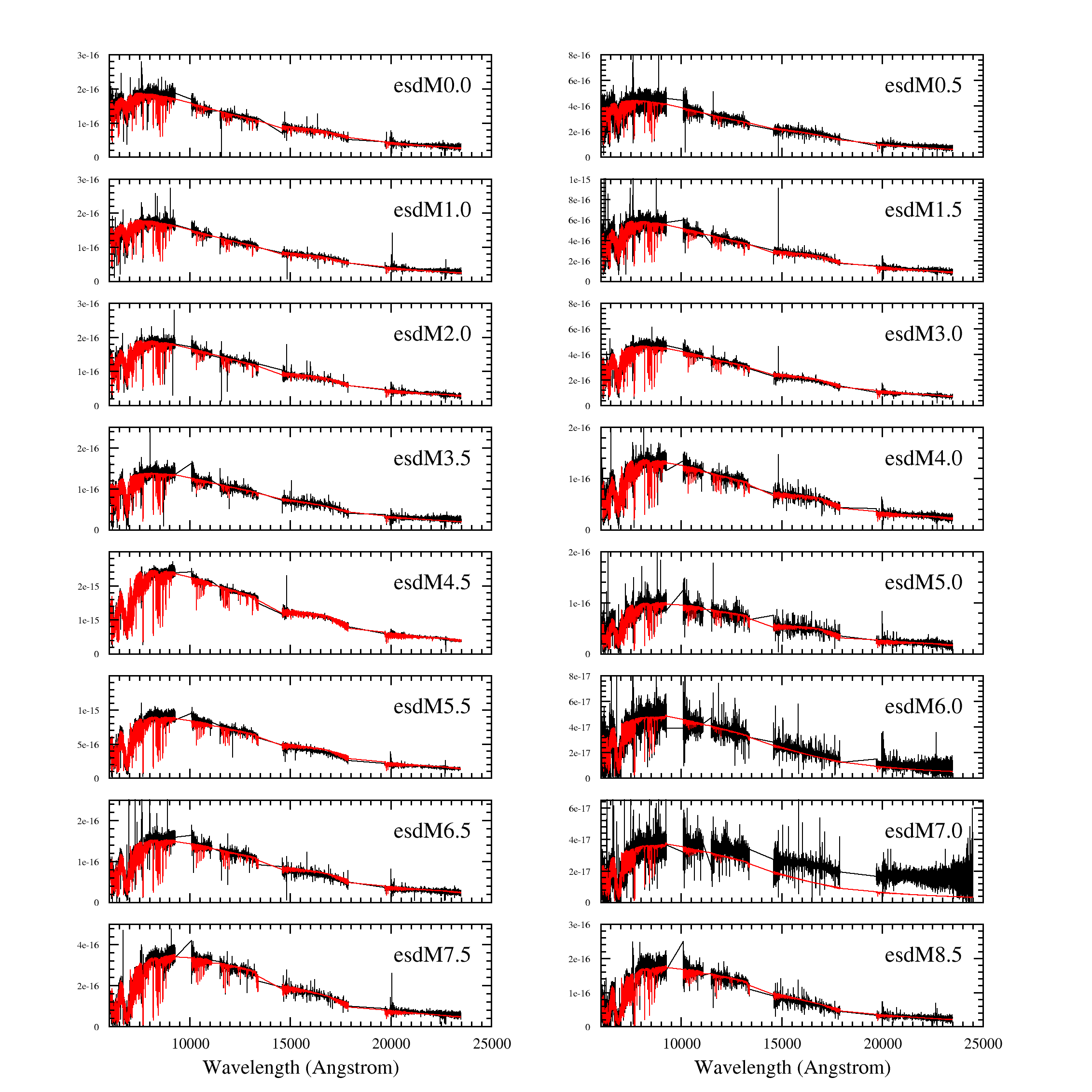}
   \caption{Same as Fig.\ \ref{fig_MplusTeff_sdM:full_XSH_model_FL_sdM} for extreme subdwarfs (esdM).
   }
   \label{fig_MplusTeff_sdM:full_XSH_model_FL_esdM}
\end{figure*}
%

%
%%%%% usdM %%%%%
%
\begin{figure*}
  \centering
  \includegraphics[width=\linewidth, angle=0]{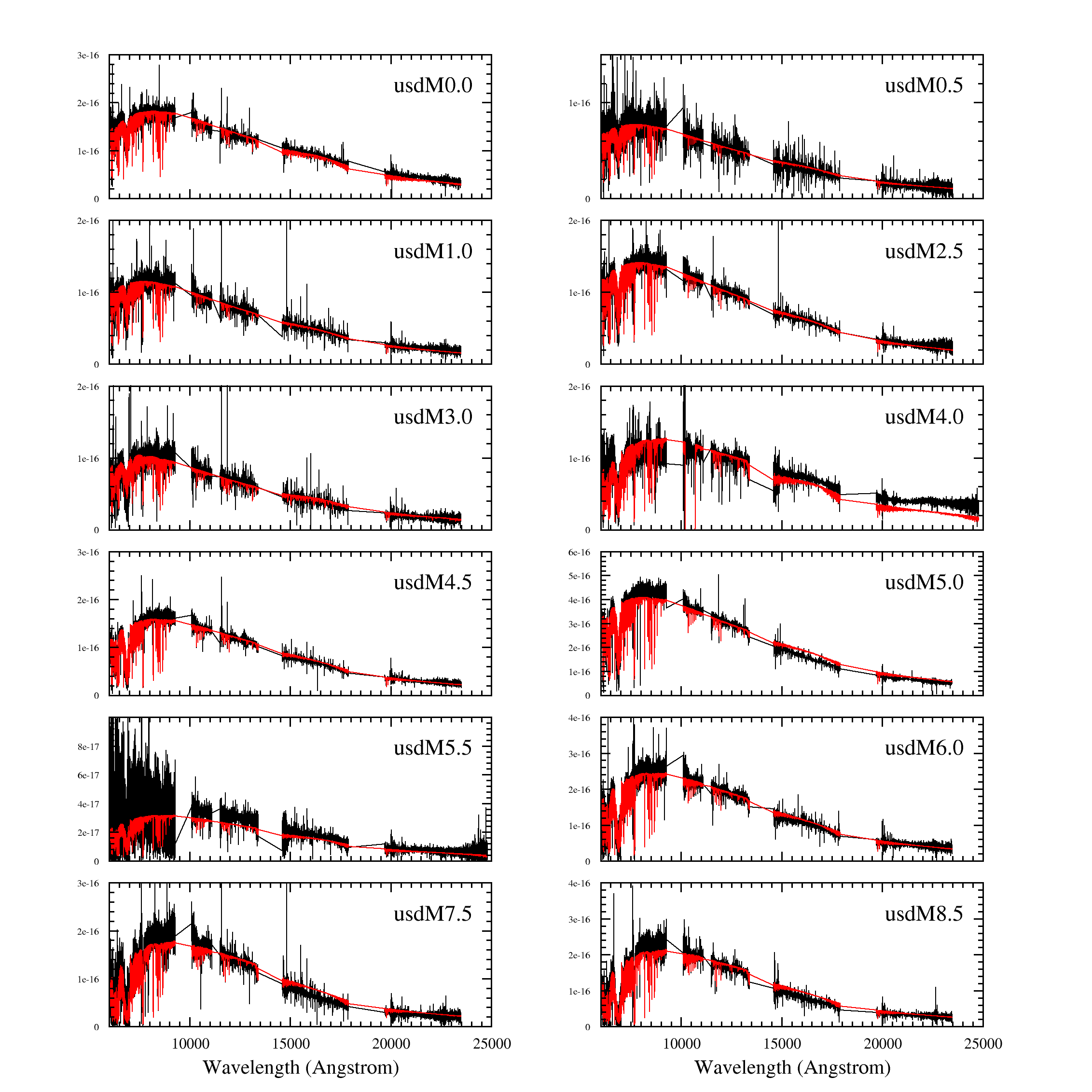}
   \caption{Same as Fig.\ \ref{fig_MplusTeff_sdM:full_XSH_model_FL_sdM} for ultra subdwarfs (usdM).
   }
   \label{fig_MplusTeff_sdM:full_XSH_model_FL_usdM}
\end{figure*}
%

%
%%%%%%%%%%%%%%%%%%%%%%%%%%%%%%%%%%%%%%%%%
%%%%% Figure: XSH spec vs model: LF %%%%%
%%%%%%%%%%%%%%%%%%%%%%%%%%%%%%%%%%%%%%%%%
%
%%%%% sdM %%%%%
%
\begin{figure*}
  \centering
  \includegraphics[width=\linewidth, angle=0]{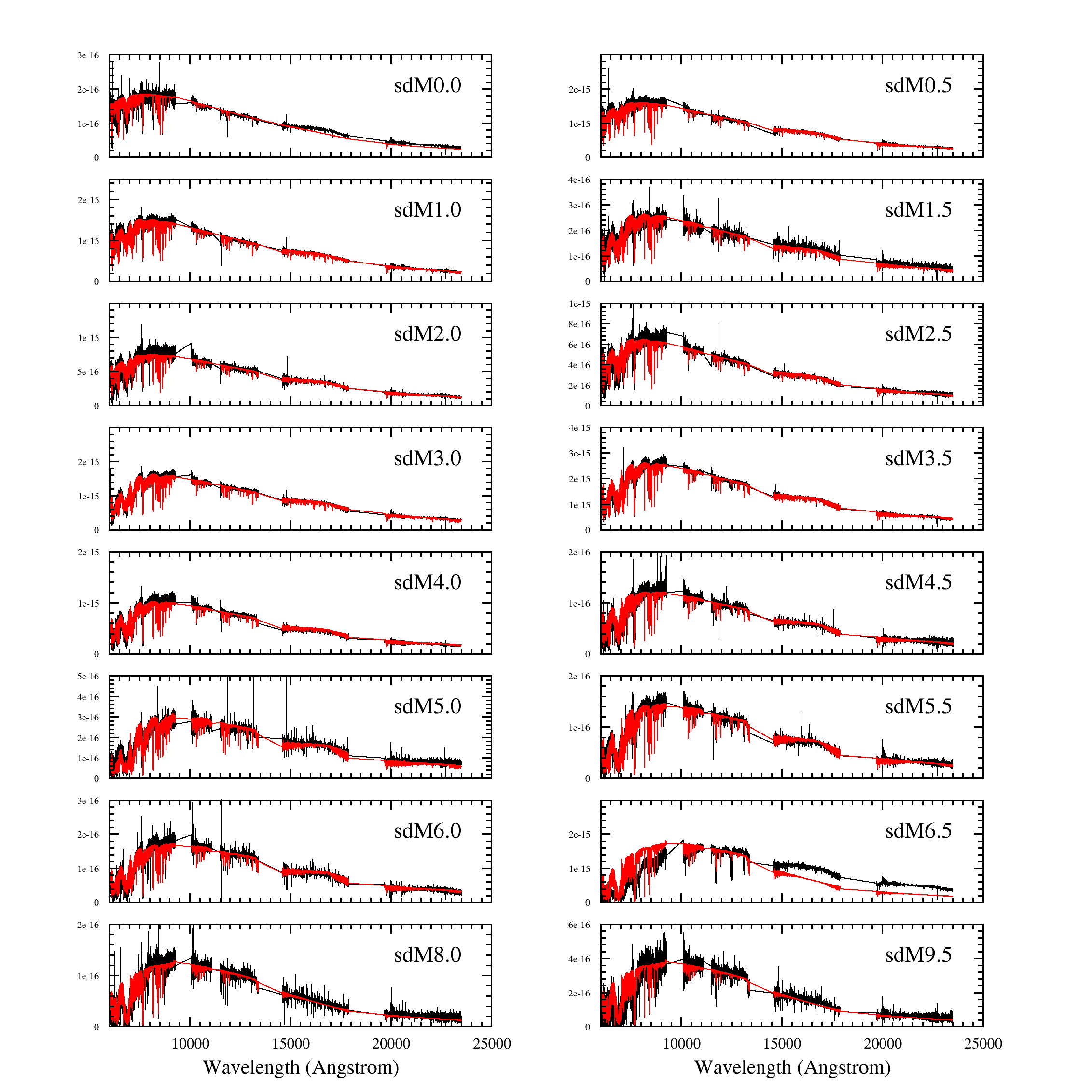}
   \caption{VLT/X-shooter UVB (450--550\,nm), VIS (550--1000 nm) and NIR (1000--2500 nm) spectra
   of sdM with spectral types from M0.0 to M9.5 (black lines) compared with the best BT-Settl
   spectra smoothed to the observed spectra. Spectral types are quoted in the top right corner
   for each subtype. The fits shown are for the LF case.
   }
   \label{fig_MplusTeff_sdM:full_XSH_model_LF_sdM}
\end{figure*}
%

%
%%%%% esdM %%%%%
%
\begin{figure*}
  \centering
  \includegraphics[width=\linewidth, angle=0]{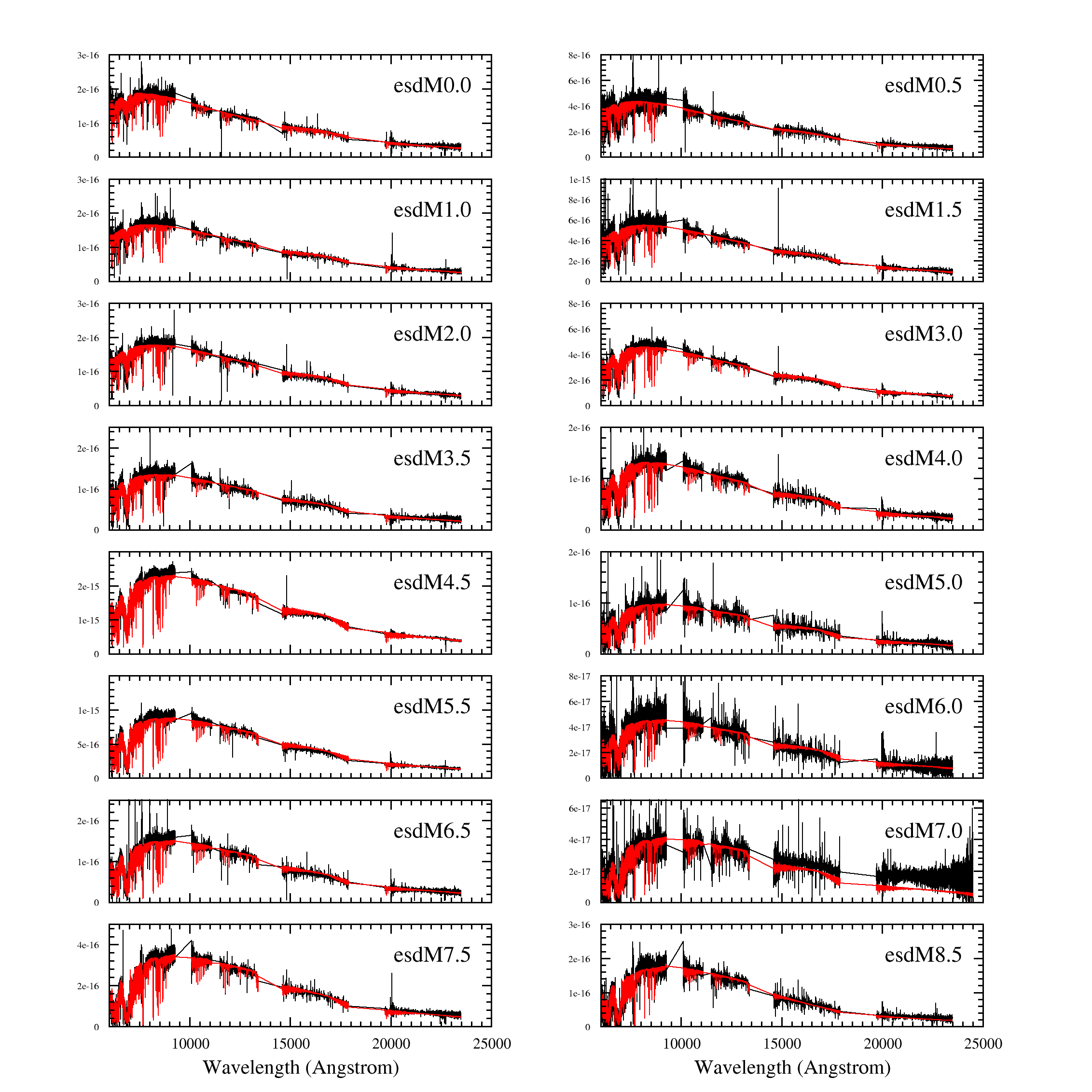}
   \caption{Same as Fig.\ \ref{fig_MplusTeff_sdM:full_XSH_model_LF_sdM} for extreme subdwarfs (esdM).
   }
   \label{fig_MplusTeff_sdM:full_XSH_model_LF_esdM}
\end{figure*}
%

%
%%%%% usdM %%%%%
%
\begin{figure*}
  \centering
  \includegraphics[width=\linewidth, angle=0]{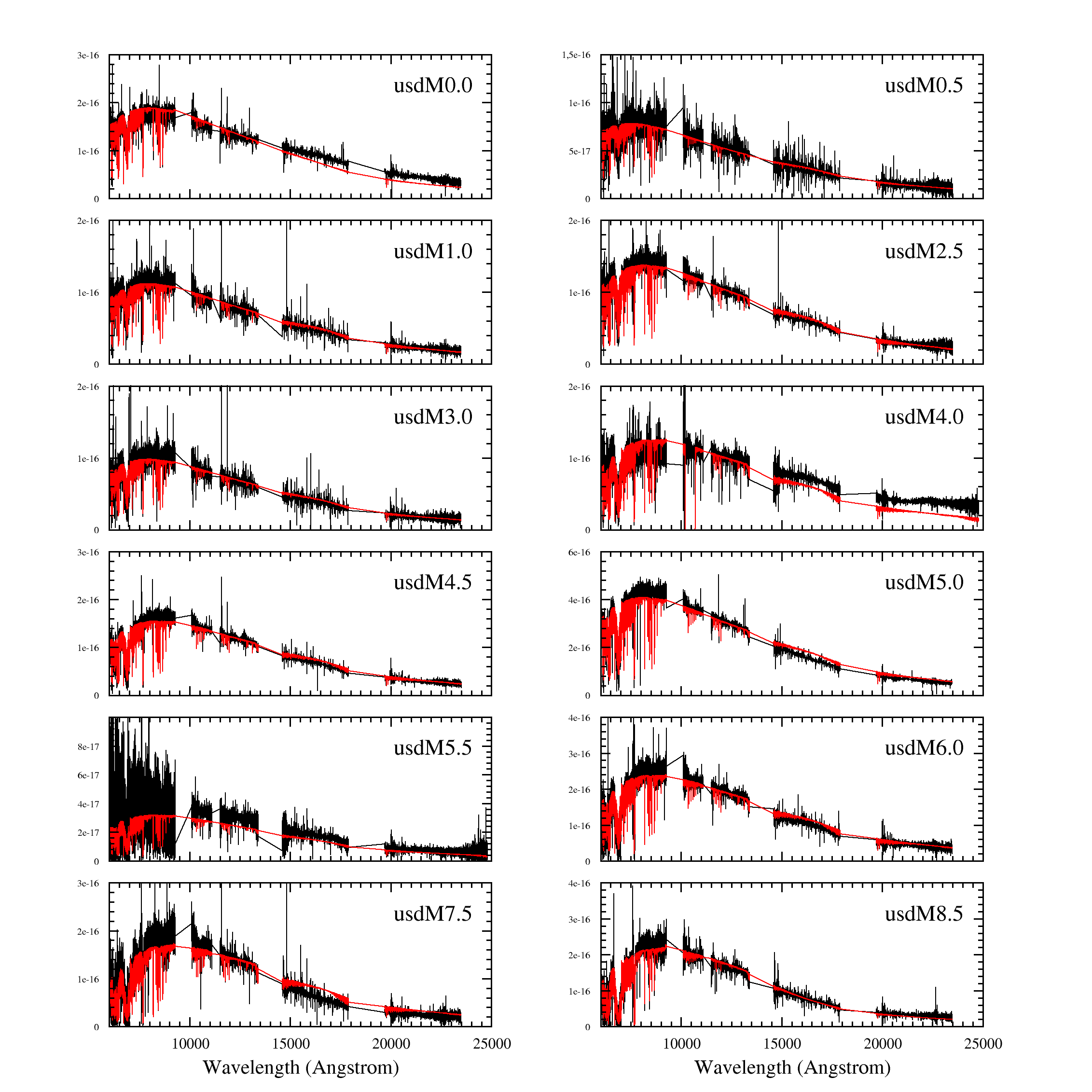}
   \caption{Same as Fig.\ \ref{fig_MplusTeff_sdM:full_XSH_model_LF_sdM} for ultra subdwarfs (usdM).
   }
   \label{fig_MplusTeff_sdM:full_XSH_model_LF_usdM}
\end{figure*}
%

%
%%%%%%%%%%%%%%%%%%%%%%%%%%%%%%%%%%%%%%%%%
%%%%% Figure: XSH spec vs model: LINES %%%%%
%%%%%%%%%%%%%%%%%%%%%%%%%%%%%%%%%%%%%%%%%
%
%%%%% sdM %%%%%
%
\begin{figure*}
  \centering
  \includegraphics[width=0.90\linewidth, angle=0]{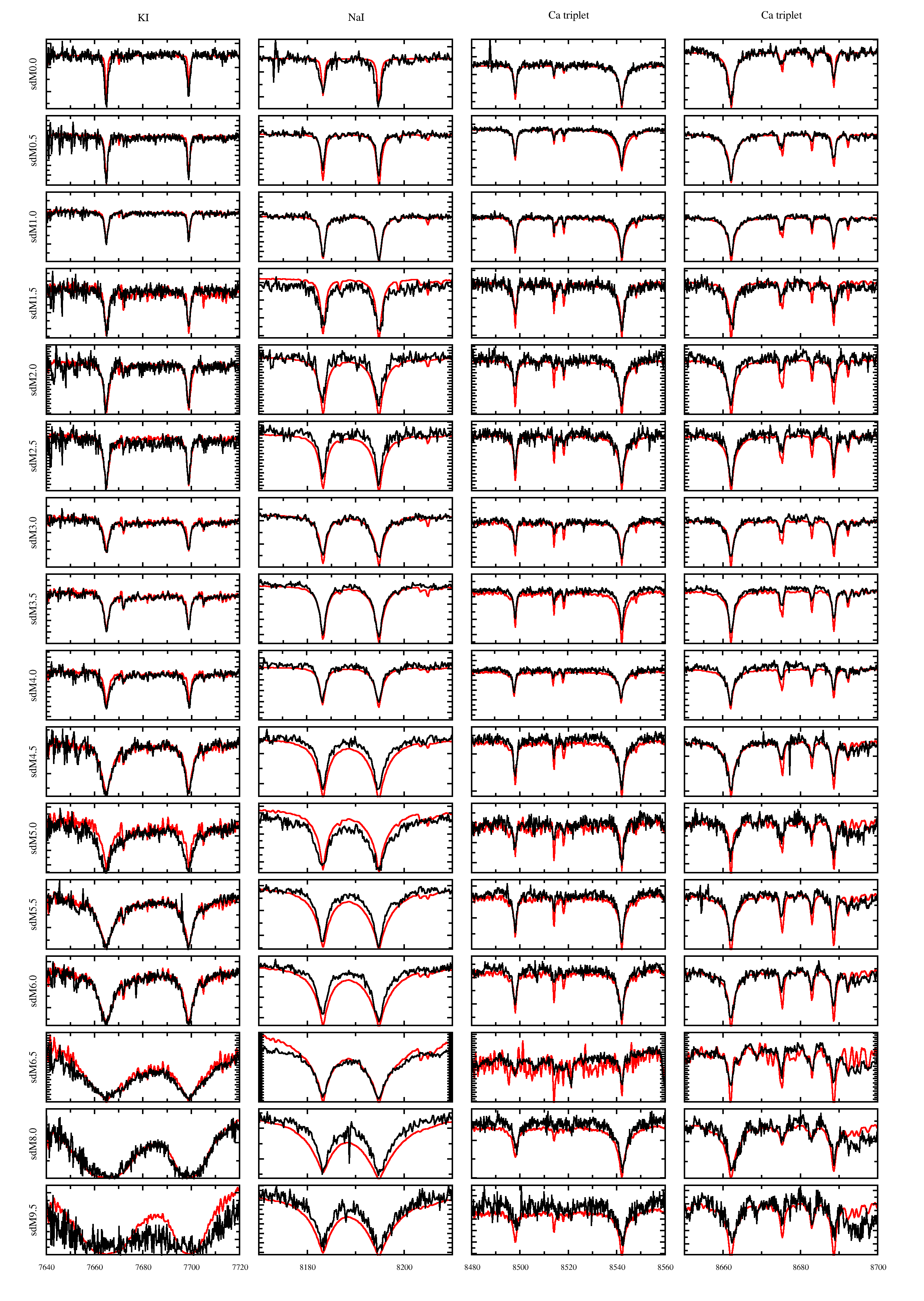}
   \caption{VLT/X-shooter UVB (450--550\,nm), VIS (550--1000 nm) and NIR (1000--2500 nm) spectra 
   of sdM with spectral types from M0.0 to M9.5 (black lines) compared with the best BT-Settl
   spectra smoothed to the observed spectra for a few strong lines (sodium, potassium, and calcium). 
   Spectral types are quoted in the top right corner for each subtype.
   }
   \label{fig_MplusTeff_sdM:full_XSH_model_lines_sdM}
\end{figure*}
%

%
%%%%% esdM %%%%%
%
\begin{figure*}
  \centering
  \includegraphics[width=0.90\linewidth, angle=0]{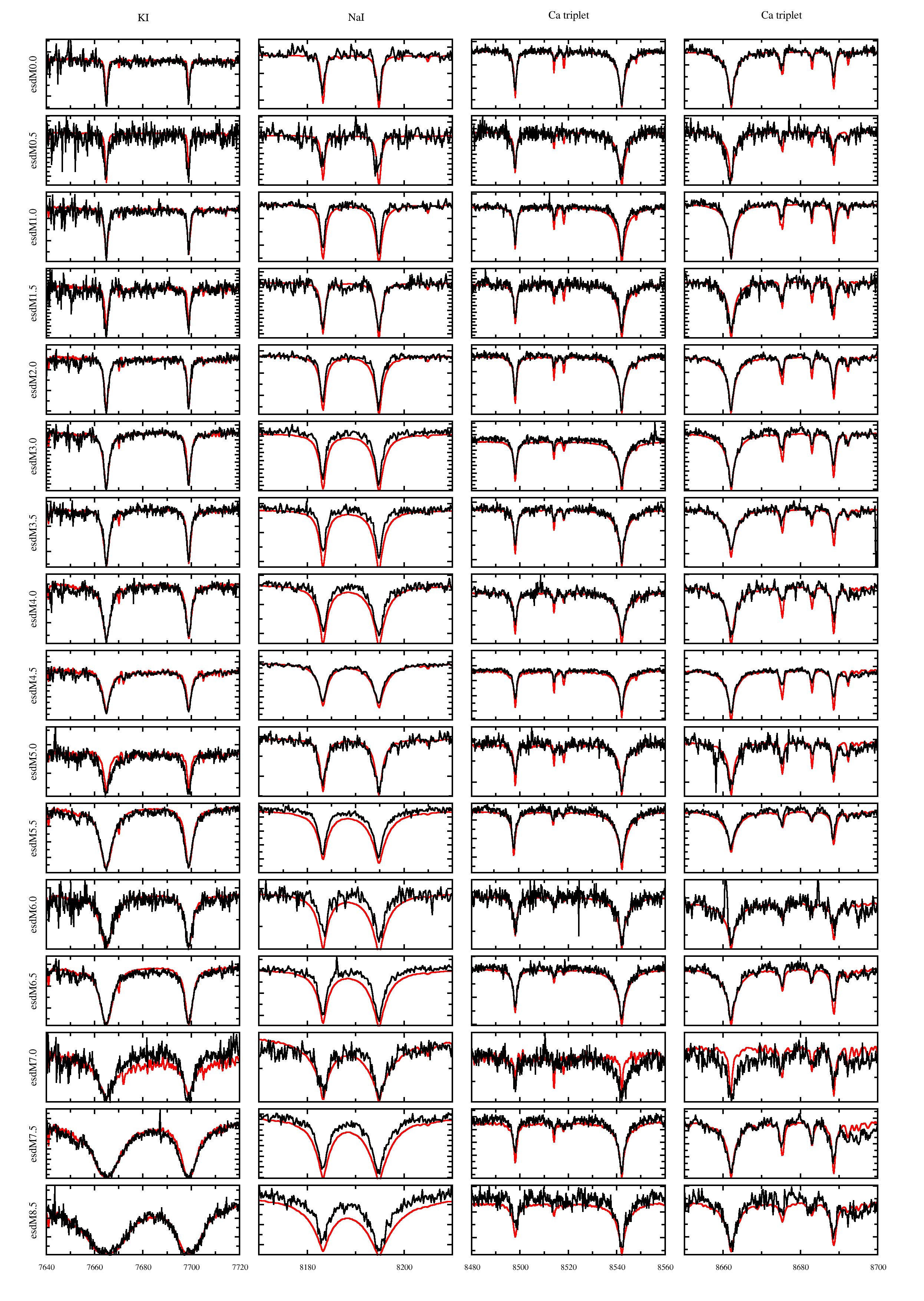}
   \caption{Same as Fig.\ \ref{fig_MplusTeff_sdM:full_XSH_model_lines_sdM} for extreme subdwarfs (esdM).
   }
   \label{fig_MplusTeff_esdM:full_XSH_model_lines_esdM}
\end{figure*}
%

%
%%%%% usdM %%%%%
%
\begin{figure*}
  \centering
  \includegraphics[width=\linewidth, angle=0]{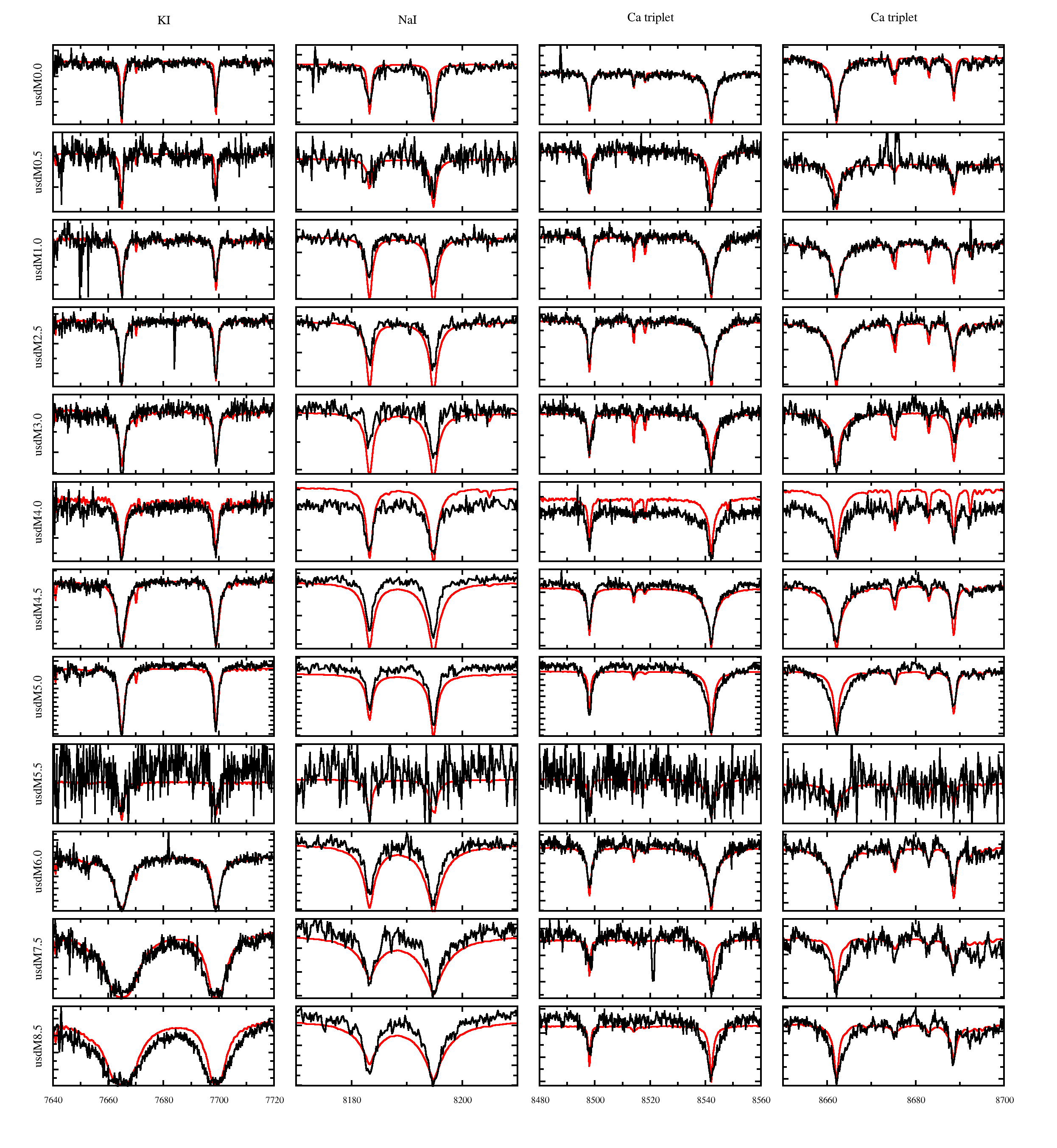}
   \caption{Same as Fig.\ \ref{fig_MplusTeff_sdM:full_XSH_model_lines_sdM} for ultra subdwarfs (usdM).
   }
   \label{fig_MplusTeff_usdM:full_XSH_model_lines_usdM}
\end{figure*}
\section{Tables with physical parameters of subdwarfs}
\label{MplusTeff_sdM:appendix_tables}
%

%
%%%%%%%%%%%%%%%%%%%%%%%%%%%%%%%%%%%%%%%%%%%%%%%%%%%%%
%%%%% Table: Physical parameters from SED fits %%%%%
%%%%%%%%%%%%%%%%%%%%%%%%%%%%%%%%%%%%%%%%%%%%%%%%%%%%%
%
\begin{table*}
\centering
\caption{Derived physical parameters for sdM, esdM, and usdM from the comparison between the observed 
X-shooter spectra and the BT-Settl synthetic spectra using the ``FF'' procedure. For each spectral subtype 
and metal class, we list the best fits with effective temperature (T$_{\rm eff}$), gravity ($\log$g), 
metallicity (M/H), chi$^{2}$ value, factor, and the name of the model.
}
\begin{tabular}{c c c c c c c c}
  \hline
  \hline
SpT  & Range  & T$_{\rm eff}$ & $\log$g  & M/H & chi$^{2}$  & Factor & Model \\
  \hline
sdM0.0 & FF & 3600 & 4.5 & -2.0 & 1.336 & 2.263e-22 & lte036.0-4.5-2.0a+0.4.BT-Settl.spec.7.dat \cr
sdM0.5 & FF & 3700 & 5.0 & -1.0 & 72.07 & 1.782e-21 & lte037.0-5.0-1.0a+0.4.BT-Settl.spec.7.dat \cr
sdM1.0 & FF & 3700 & 5.0 & -0.5 & 67.22 & 1.609e-21 & lte037.0-5.0-0.5a+0.2.BT-Settl.spec.7.dat \cr
sdM1.5 & FF & 3600 & 4.5 & -0.0 & 66.4 & 3.168e-22 & lte036.0-4.5-0.0a+0.0.BT-Settl.spec.7.dat \cr
sdM2.0 & FF & 3600 & 5.5 & -0.5 & 46.01 & 9.281e-22 & lte036.0-5.5-0.5a+0.2.BT-Settl.spec.7.dat \cr
sdM2.5 & FF & 3600 & 5.5 & -0.5 & 35.09 & 7.689e-22 & lte036.0-5.5-0.5a+0.2.BT-Settl.spec.7.dat \cr
sdM3.0 & FF & 3500 & 5.0 & -0.0 & 88.51 & 2.208e-21 & lte035.0-5.0-0.0a+0.0.BT-Settl.spec.7.dat \cr
sdM3.5 & FF & 3400 & 5.0 & -0.5 & 161.6 & 3.916e-21 & lte034.0-5.0-0.5a+0.2.BT-Settl.spec.7.dat \cr
sdM4.0 & FF & 3400 & 5.0 & -1.0 & 58.22 & 1.54e-21 & lte034.0-5.0-1.0a+0.4.BT-Settl.spec.7.dat \cr
sdM4.5 & FF & 3300 & 5.5 & -1.0 & 59.07 & 2.072e-22 & lte033.0-5.5-1.0a+0.4.BT-Settl.spec.7.dat \cr
sdM5.0 & FF & 3200 & 5.0 & -0.0 & 62.52 & 5.826e-22 & lte032.0-5.0-0.0a+0.0.BT-Settl.spec.7.dat \cr
sdM5.5 & FF & 3200 & 5.5 & -1.0 & 60.94 & 2.764e-22 & lte032.0-5.5-1.0a+0.4.BT-Settl.spec.7.dat \cr
sdM6.0 & FF & 3200 & 5.5 & -1.0 & 52.62 & 3.327e-22 & lte032.0-5.5-1.0a+0.4.BT-Settl.spec.7.dat \cr
sdM6.5 & FF & 2900 & 5.5 & -0.0 & 828.4 & 4.589e-21 & lte029.0-5.5-0.0a+0.0.BT-Settl.spec.7.dat \cr
sdM8.0 & FF & 2900 & 5.5 & -2.0 & 49.54 & 3.094e-22 & lte029.0-5.5-2.0a+0.4.BT-Settl.spec.7.dat \cr
sdM9.5 & FF & 2800 & 5.0 & -2.0 & 26.84 & 1.071e-21 & lte028.0-5.0-2.0a+0.4.BT-Settl.spec.7.dat \cr
 \hline
esdM0.0 & FF & 3800 & 5.0 & -1.0 & 46.36 & 1.789e-22 & lte038.0-5.0-1.0a+0.4.BT-Settl.spec.7.dat \cr
esdM0.5 & FF & 3700 & 5.0 & -1.5 & 28.12 & 4.847e-22 & lte037.0-5.0-1.5a+0.4.BT-Settl.spec.7.dat \cr
esdM1.0 & FF & 3800 & 5.5 & -0.5 & 50.66 & 1.683e-22 & lte038.0-5.5-0.5a+0.2.BT-Settl.spec.7.dat \cr
esdM1.5 & FF & 3600 & 5.0 & -1.0 & 24.61 & 6.996e-22 & lte036.0-5.0-1.0a+0.4.BT-Settl.spec.7.dat \cr
esdM2.0 & FF & 3600 & 5.5 & -1.0 & 54.63 & 2.229e-22 & lte036.0-5.5-1.0a+0.4.BT-Settl.spec.7.dat \cr
esdM3.0 & FF & 3500 & 5.5 & -1.5 & 82.5 & 6.294e-22 & lte035.0-5.5-1.5a+0.4.BT-Settl.spec.7.dat \cr
esdM3.5 & FF & 3500 & 5.5 & -1.5 & 38 & 1.883e-22 & lte035.0-5.5-1.5a+0.4.BT-Settl.spec.7.dat \cr
esdM4.0 & FF & 3400 & 5.5 & -1.5 & 53.24 & 2.04e-22 & lte034.0-5.5-1.5a+0.4.BT-Settl.spec.7.dat \cr
esdM4.5 & FF & 3400 & 5.5 & -1.0 & 260.6 & 3.594e-21 & lte034.0-5.5-1.0a+0.4.BT-Settl.spec.7.dat \cr
esdM5.0 & FF & 3200 & 4.5 & -1.5 & 39.72 & 1.952e-22 & lte032.0-4.5-1.5a+0.4.BT-Settl.spec.7.dat \cr
esdM5.5 & FF & 3300 & 5.5 & -2.0 & 87.47 & 1.533e-21 & lte033.0-5.5-2.0a+0.4.BT-Settl.spec.7.dat \cr
esdM6.0 & FF & 3300 & 5.5 & -2.0 & 27.35 & 8.065e-23 & lte033.0-5.5-2.0a+0.4.BT-Settl.spec.7.dat \cr
esdM6.5 & FF & 3300 & 5.5 & -2.0 & 84.99 & 2.61e-22 & lte033.0-5.5-2.0a+0.4.BT-Settl.spec.7.dat \cr
esdM7.0 & FF & 3200 & 5.5 & -0.0 & 8.431 & 8.349e-23 & lte032.0-5.5-0.0a+0.0.BT-Settl.spec.7.dat \cr
esdM7.5 & FF & 3000 & 5.0 & -2.0 & 75.73 & 8.239e-22 & lte030.0-5.0-2.0a+0.4.BT-Settl.spec.7.dat \cr
esdM8.5 & FF & 3000 & 5.5 & -2.0 & 46.61 & 3.916e-22 & lte030.0-5.5-2.0a+0.4.BT-Settl.spec.7.dat \cr
  \hline
usdM0.0 & FF & 3500 & 4.5 & -2.0 & 81.52 & 2.637e-22 & lte035.0-4.5-2.0a+0.4.BT-Settl.spec.7.dat \cr
usdM0.5 & FF & 3700 & 5.5 & -2.5 & 35.49 & 8.444e-23 & lte037.0-5.5-2.5a+0.4.BT-Settl.spec.7.dat \cr
usdM1.0 & FF & 3700 & 5.5 & -1.0 & 34.9 & 1.245e-22 & lte037.0-5.5-1.0a+0.4.BT-Settl.spec.7.dat \cr
usdM2.5 & FF & 3600 & 5.5 & -1.5 & 40.94 & 1.724e-22 & lte036.0-5.5-1.5a+0.4.BT-Settl.spec.7.dat \cr
usdM3.0 & FF & 3800 & 6.0 & -0.5 & 40.92 & 9.605e-23 & lte038.0-6.0-0.5a+0.2.BT-Settl.spec.7.dat \cr
usdM4.0 & FF & 3200 & 4.5 & -1.5 & 8.887 & 2.545e-22 & lte032.0-4.5-1.5a+0.4.BT-Settl.spec.7.dat \cr
usdM4.5 & FF & 3400 & 5.5 & -2.0 & 79.52 & 2.428e-22 & lte034.0-5.5-2.0a+0.4.BT-Settl.spec.7.dat \cr
usdM5.0 & FF & 3500 & 5.5 & -2.5 & 128.5 & 5.667e-22 & lte035.0-5.5-2.5a+0.4.BT-Settl.spec.7.dat \cr
usdM5.5 & FF & 3300 & 4.5 & -2.0 & 12.52 & 5.669e-23 & lte033.0-4.5-2.0a+0.4.BT-Settl.spec.7.dat \cr
usdM6.0 & FF & 3300 & 5.5 & -2.0 & 40.62 & 4.176e-22 & lte033.0-5.5-2.0a+0.4.BT-Settl.spec.7.dat \cr
usdM7.5 & FF & 3100 & 5.5 & -2.5 & 25.7 & 3.749e-22 & lte031.0-5.5-2.5a+0.4.BT-Settl.spec.7.dat \cr
usdM8.5 & FF & 3100 & 5.5 & -2.5 & 49.69 & 4.503e-22 & lte031.0-5.5-2.5a+0.4.BT-Settl.spec.7.dat \cr
  \hline
  \label{tab_MplusTeff_sdM:table_best_fits_FF}
\end{tabular}
\end{table*}
%

%
%%%%% LL option %%%%%
%
\begin{table*}
\centering
\caption{Derived physical parameters for sdM, esdM, and usdM from the comparison between the observed 
X-shooter spectra and the BT-Settl synthetic spectra using the ``LL'' procedure.
}
\begin{tabular}{c c c c c c c c}
  \hline
  \hline
SpT  & Range  & T$_{\rm eff}$ & $\log$g  & M/H & chi$^{2}$  & Factor & Model \\
\hline
sdM0.0 & LL & 3600 & 5.5 & -2.5 & 1.274 & 2.182e-22 & lte036.0-5.5-2.5a+0.4.BT-Settl.spec.7.dat \cr
sdM0.5 & LL & 3500 & 4.5 & -1.5 & 65.2 & 2.303e-21 & lte035.0-4.5-1.5a+0.4.BT-Settl.spec.7.dat \cr
sdM1.0 & LL & 3600 & 5.0 & -0.5 & 44.52 & 1.843e-21 & lte036.0-5.0-0.5a+0.2.BT-Settl.spec.7.dat \cr
sdM1.5 & LL & 3600 & 5.0 & -0.5 & 53.5 & 2.944e-22 & lte036.0-5.0-0.5a+0.2.BT-Settl.spec.7.dat \cr
sdM2.0 & LL & 3300 & 4.5 & -1.5 & 54.06 & 1.418e-21 & lte033.0-4.5-1.5a+0.4.BT-Settl.spec.7.dat \cr
sdM2.5 & LL & 3400 & 5.0 & -1.0 & 33.27 & 1.013e-21 & lte034.0-5.0-1.0a+0.4.BT-Settl.spec.7.dat \cr
sdM3.0 & LL & 3500 & 5.0 & -0.0 & 86.06 & 2.228e-21 & lte035.0-5.0-0.0a+0.0.BT-Settl.spec.7.dat \cr
sdM3.5 & LL & 3300 & 5.0 & -1.0 & 100 & 4.559e-21 & lte033.0-5.0-1.0a+0.4.BT-Settl.spec.7.dat \cr
sdM4.0 & LL & 3300 & 5.0 & -1.0 & 43 & 1.854e-21 & lte033.0-5.0-1.0a+0.4.BT-Settl.spec.7.dat \cr
sdM4.5 & LL & 3200 & 5.0 & -1.5 & 64.01 & 2.467e-22 & lte032.0-5.0-1.5a+0.4.BT-Settl.spec.7.dat \cr
sdM5.0 & LL & 3200 & 5.5 & -0.5 & 64.17 & 5.346e-22 & lte032.0-5.5-0.5a+0.2.BT-Settl.spec.7.dat \cr
sdM5.5 & LL & 3100 & 5.0 & -1.5 & 62.38 & 3.231e-22 & lte031.0-5.0-1.5a+0.4.BT-Settl.spec.7.dat \cr
sdM6.0 & LL & 3000 & 4.5 & -1.5 & 68.92 & 4.55e-22 & lte030.0-4.5-1.5a+0.4.BT-Settl.spec.7.dat \cr
sdM6.5 & LL & 2600 & 4.5 & -2.5 & 360.4 & 4.65e-21 & lte026.0-4.5-2.5a+0.4.BT-Settl.spec.7.dat \cr
sdM8.0 & LL & 2800 & 5.0 & -2.5 & 68.93 & 3.399e-22 & lte028.0-5.0-2.5a+0.4.BT-Settl.spec.7.dat \cr
sdM9.5 & LL & 2700 & 5.0 & -2.5 & 33.25 & 1.083e-21 & lte027.0-5.0-2.5a+0.4.BT-Settl.spec.7.dat \cr
  \hline
esdM0.0 & LL & 3800 & 5.0 & -1.0 & 52.51 & 1.807e-22 & lte038.0-5.0-1.0a+0.4.BT-Settl.spec.7.dat \cr
esdM0.5 & LL & 3600 & 5.0 & -2.0 & 36.42 & 5.423e-22 & lte036.0-5.0-2.0a+0.4.BT-Settl.spec.7.dat \cr
esdM1.0 & LL & 3500 & 4.5 & -2.0 & 43.43 & 2.508e-22 & lte035.0-4.5-2.0a+0.4.BT-Settl.spec.7.dat \cr
esdM1.5 & LL & 3400 & 4.5 & -2.0 & 32.73 & 9.175e-22 & lte034.0-4.5-2.0a+0.4.BT-Settl.spec.7.dat \cr
esdM2.0 & LL & 3300 & 4.5 & -2.0 & 41.6 & 3.457e-22 & lte033.0-4.5-2.0a+0.4.BT-Settl.spec.7.dat \cr
esdM3.0 & LL & 3400 & 5.0 & -1.5 & 54.27 & 7.476e-22 & lte034.0-5.0-1.5a+0.4.BT-Settl.spec.7.dat \cr
esdM3.5 & LL & 3400 & 5.0 & -2.0 & 30.57 & 2.211e-22 & lte034.0-5.0-2.0a+0.4.BT-Settl.spec.7.dat \cr
esdM4.0 & LL & 3300 & 5.0 & -1.5 & 60.87 & 2.371e-22 & lte033.0-5.0-1.5a+0.4.BT-Settl.spec.7.dat \cr
esdM4.5 & LL & 3100 & 4.5 & -2.0 & 188.1 & 5.634e-21 & lte031.0-4.5-2.0a+0.4.BT-Settl.spec.7.dat \cr
esdM5.0 & LL & 3100 & 4.5 & -2.0 & 36.16 & 2.286e-22 & lte031.0-4.5-2.0a+0.4.BT-Settl.spec.7.dat \cr
esdM5.5 & LL & 3200 & 5.0 & -2.0 & 65.67 & 1.831e-21 & lte032.0-5.0-2.0a+0.4.BT-Settl.spec.7.dat \cr
esdM6.0 & LL & 3100 & 4.5 & -2.0 & 29.34 & 1.116e-22 & lte031.0-4.5-2.0a+0.4.BT-Settl.spec.7.dat \cr
esdM6.5 & LL & 3200 & 5.0 & -2.0 & 62.31 & 3.104e-22 & lte032.0-5.0-2.0a+0.4.BT-Settl.spec.7.dat \cr
esdM7.0 & LL & 3000 & 4.5 & -2.0 & 2.347 & 9.268e-23 & lte030.0-4.5-2.0a+0.4.BT-Settl.spec.7.dat \cr
esdM7.5 & LL & 3000 & 5.0 & -2.0 & 93.43 & 8.48e-22 & lte030.0-5.0-2.0a+0.4.BT-Settl.spec.7.dat \cr
esdM8.5 & LL & 2900 & 5.0 & -2.5 & 54.67 & 4.499e-22 & lte029.0-5.0-2.5a+0.4.BT-Settl.spec.7.dat \cr
  \hline
usdM0.0 & LL & 3600 & 5.5 & -2.5 & 52.74 & 2.186e-22 & lte036.0-5.5-2.5a+0.4.BT-Settl.spec.7.dat \cr
usdM0.5 & LL & 3700 & 5.5 & -2.0 & 52.29 & 8.588e-23 & lte037.0-5.5-2.0a+0.4.BT-Settl.spec.7.dat \cr
usdM1.0 & LL & 3600 & 5.0 & -2.0 & 38.42 & 1.449e-22 & lte036.0-5.0-2.0a+0.4.BT-Settl.spec.7.dat \cr
usdM2.5 & LL & 3500 & 5.0 & -2.0 & 35.32 & 1.999e-22 & lte035.0-5.0-2.0a+0.4.BT-Settl.spec.7.dat \cr
usdM3.0 & LL & 3600 & 5.5 & -2.0 & 38.99 & 1.27e-22 & lte036.0-5.5-2.0a+0.4.BT-Settl.spec.7.dat \cr
usdM4.0 & LL & 3400 & 5.0 & -2.0 & 0.7288 & 1.663e-22 & lte034.0-5.0-2.0a+0.4.BT-Settl.spec.7.dat \cr
usdM4.5 & LL & 3300 & 5.0 & -2.0 & 81.28 & 2.949e-22 & lte033.0-5.0-2.0a+0.4.BT-Settl.spec.7.dat \cr
usdM5.0 & LL & 3500 & 5.5 & -2.0 & 91.84 & 5.981e-22 & lte035.0-5.5-2.0a+0.4.BT-Settl.spec.7.dat \cr
usdM5.5 & LL & 3500 & 5.5 & -2.0 & 11.58 & 3.992e-23 & lte035.0-5.5-2.0a+0.4.BT-Settl.spec.7.dat \cr
usdM6.0 & LL & 3200 & 5.0 & -2.0 & 44.5 & 5.027e-22 & lte032.0-5.0-2.0a+0.4.BT-Settl.spec.7.dat \cr
usdM7.5 & LL & 3000 & 5.0 & -2.0 & 39.86 & 4.725e-22 & lte030.0-5.0-2.0a+0.4.BT-Settl.spec.7.dat \cr
usdM8.5 & LL & 3000 & 5.5 & -2.5 & 60.58 & 5.247e-22 & lte030.0-5.5-2.5a+0.4.BT-Settl.spec.7.dat \cr
  \hline
  \label{tab_MplusTeff_sdM:table_best_fits_LL}
\end{tabular}
\end{table*}
%

%
%%%%% FL option %%%%%
%
\begin{table*}
\centering
\caption{Derived physical parameters for sdM, esdM, and usdM from the comparison between the observed 
X-shooter spectra and the BT-Settl synthetic spectra using the ``FL'' procedure. For each spectral subtype 
and metal class, we list the best fits with effective temperature (T$_{\rm eff}$), gravity ($\log$g), 
metallicity (M/H), chi$^{2}$ value, factor, and the name of the model.
}
\begin{tabular}{c c c c c c c c}
  \hline
  \hline
SpT  & Range  & T$_{\rm eff}$ & $\log$g  & M/H & chi$^{2}$  & Factor & Model \\
  \hline
sdM0.0 & FL & 3600 & 5.5 & -2.5 & 1.274 & 2.182e-22 & lte036.0-5.5-2.5a+0.4.BT-Settl.spec.7.dat \cr
sdM0.5 & FL & 3700 & 5.0 & -0.5 & 67.14 & 1.791e-21 & lte037.0-5.0-0.5a+0.2.BT-Settl.spec.7.dat \cr
sdM1.0 & FL & 3700 & 5.0 & -0.5 & 45.51 & 1.627e-21 & lte037.0-5.0-0.5a+0.2.BT-Settl.spec.7.dat \cr
sdM1.5 & FL & 3600 & 5.0 & -0.5 & 53.5 & 2.944e-22 & lte036.0-5.0-0.5a+0.2.BT-Settl.spec.7.dat \cr
sdM2.0 & FL & 3600 & 5.0 & -0.0 & 60.82 & 9.902e-22 & lte036.0-5.0-0.0a+0.0.BT-Settl.spec.7.dat \cr
sdM2.5 & FL & 3600 & 5.0 & -0.0 & 35.7 & 8.194e-22 & lte036.0-5.0-0.0a+0.0.BT-Settl.spec.7.dat \cr
sdM3.0 & FL & 3500 & 5.0 & -0.0 & 86.06 & 2.228e-21 & lte035.0-5.0-0.0a+0.0.BT-Settl.spec.7.dat \cr
sdM3.5 & FL & 3400 & 5.0 & -0.0 & 102.6 & 4.24e-21 & lte034.0-5.0-0.0a+0.0.BT-Settl.spec.7.dat \cr
sdM4.0 & FL & 3400 & 5.0 & -0.5 & 49.18 & 1.64e-21 & lte034.0-5.0-0.5a+0.2.BT-Settl.spec.7.dat \cr
sdM4.5 & FL & 3300 & 5.5 & -1.0 & 65.18 & 2.143e-22 & lte033.0-5.5-1.0a+0.4.BT-Settl.spec.7.dat \cr
sdM5.0 & FL & 3200 & 5.5 & -0.5 & 64.17 & 5.346e-22 & lte032.0-5.5-0.5a+0.2.BT-Settl.spec.7.dat \cr
sdM5.5 & FL & 3200 & 5.5 & -1.0 & 64.82 & 2.837e-22 & lte032.0-5.5-1.0a+0.4.BT-Settl.spec.7.dat \cr
sdM6.0 & FL & 3200 & 5.5 & -1.0 & 71.23 & 3.416e-22 & lte032.0-5.5-1.0a+0.4.BT-Settl.spec.7.dat \cr
sdM6.5 & FL & 2900 & 5.5 & -1.0 & 457.5 & 3.619e-21 & lte029.0-5.5-1.0a+0.4.BT-Settl.spec.7.dat \cr
sdM8.0 & FL & 2900 & 5.5 & -2.0 & 70.63 & 3.246e-22 & lte029.0-5.5-2.0a+0.4.BT-Settl.spec.7.dat \cr
sdM9.5 & FL & 2800 & 5.5 & -2.0 & 36.7 & 1.048e-21 & lte028.0-5.5-2.0a+0.4.BT-Settl.spec.7.dat \cr
  \hline
esdM0.0 & FL & 3800 & 5.0 & -1.0 & 52.51 & 1.807e-22 & lte038.0-5.0-1.0a+0.4.BT-Settl.spec.7.dat \cr
esdM0.5 & FL & 3700 & 5.5 & -2.0 & 36.68 & 4.752e-22 & lte037.0-5.5-2.0a+0.4.BT-Settl.spec.7.dat \cr
esdM1.0 & FL & 3800 & 5.5 & -1.0 & 51.58 & 1.689e-22 & lte038.0-5.5-1.0a+0.4.BT-Settl.spec.7.dat \cr
esdM1.5 & FL & 3600 & 5.5 & -1.0 & 35.37 & 6.833e-22 & lte036.0-5.5-1.0a+0.4.BT-Settl.spec.7.dat \cr
esdM2.0 & FL & 3600 & 5.5 & -1.0 & 48.16 & 2.251e-22 & lte036.0-5.5-1.0a+0.4.BT-Settl.spec.7.dat \cr
esdM3.0 & FL & 3500 & 5.5 & -1.5 & 61.37 & 6.495e-22 & lte035.0-5.5-1.5a+0.4.BT-Settl.spec.7.dat \cr
esdM3.5 & FL & 3500 & 5.5 & -2.0 & 34.19 & 1.914e-22 & lte035.0-5.5-2.0a+0.4.BT-Settl.spec.7.dat \cr
esdM4.0 & FL & 3400 & 5.5 & -1.0 & 67.85 & 2.034e-22 & lte034.0-5.5-1.0a+0.4.BT-Settl.spec.7.dat \cr
esdM4.5 & FL & 3400 & 5.5 & -1.0 & 211.3 & 3.624e-21 & lte034.0-5.5-1.0a+0.4.BT-Settl.spec.7.dat \cr
esdM5.0 & FL & 3200 & 5.0 & -1.5 & 40.53 & 1.967e-22 & lte032.0-5.0-1.5a+0.4.BT-Settl.spec.7.dat \cr
esdM5.5 & FL & 3300 & 5.5 & -1.5 & 74.79 & 1.598e-21 & lte033.0-5.5-1.5a+0.4.BT-Settl.spec.7.dat \cr
esdM6.0 & FL & 3300 & 5.5 & -2.5 & 31.14 & 8.05e-23 & lte033.0-5.5-2.5a+0.4.BT-Settl.spec.7.dat \cr
esdM6.5 & FL & 3300 & 5.5 & -1.5 & 72.68 & 2.71e-22 & lte033.0-5.5-1.5a+0.4.BT-Settl.spec.7.dat \cr
esdM7.0 & FL & 3200 & 5.5 & -2.5 & 2.51 & 6.41e-23 & lte032.0-5.5-2.5a+0.4.BT-Settl.spec.7.dat \cr
esdM7.5 & FL & 3000 & 5.0 & -2.0 & 93.43 & 8.48e-22 & lte030.0-5.0-2.0a+0.4.BT-Settl.spec.7.dat \cr
esdM8.5 & FL & 3000 & 5.5 & -2.0 & 54.82 & 4.189e-22 & lte030.0-5.5-2.0a+0.4.BT-Settl.spec.7.dat \cr
  \hline
usdM0.0 & FL & 3500 & 4.5 & -2.0 & 54.08 & 2.592e-22 & lte035.0-4.5-2.0a+0.4.BT-Settl.spec.7.dat \cr
usdM0.5 & FL & 3700 & 5.5 & -2.0 & 52.29 & 8.588e-23 & lte037.0-5.5-2.0a+0.4.BT-Settl.spec.7.dat \cr
usdM1.0 & FL & 3700 & 5.5 & -2.0 & 38.67 & 1.27e-22 & lte037.0-5.5-2.0a+0.4.BT-Settl.spec.7.dat \cr
usdM2.5 & FL & 3600 & 5.5 & -2.0 & 35.69 & 1.745e-22 & lte036.0-5.5-2.0a+0.4.BT-Settl.spec.7.dat \cr
usdM3.0 & FL & 3800 & 5.5 & -0.5 & 46.05 & 1.005e-22 & lte038.0-5.5-0.5a+0.2.BT-Settl.spec.7.dat \cr
usdM4.0 & FL & 3200 & 4.5 & -2.0 & 0.772 & 2.22e-22 & lte032.0-4.5-2.0a+0.4.BT-Settl.spec.7.dat \cr
usdM4.5 & FL & 3400 & 5.5 & -2.0 & 85.36 & 2.539e-22 & lte034.0-5.5-2.0a+0.4.BT-Settl.spec.7.dat \cr
usdM5.0 & FL & 3500 & 5.5 & -2.0 & 91.84 & 5.981e-22 & lte035.0-5.5-2.0a+0.4.BT-Settl.spec.7.dat \cr
usdM5.5 & FL & 3300 & 5.0 & -2.0 & 11.84 & 5.247e-23 & lte033.0-5.0-2.0a+0.4.BT-Settl.spec.7.dat \cr
usdM6.0 & FL & 3300 & 5.5 & -2.0 & 47.7 & 4.314e-22 & lte033.0-5.5-2.0a+0.4.BT-Settl.spec.7.dat \cr
usdM7.5 & FL & 3100 & 5.5 & -2.0 & 40.37 & 4.046e-22 & lte031.0-5.5-2.0a+0.4.BT-Settl.spec.7.dat \cr
usdM8.5 & FL & 3100 & 5.5 & -2.0 & 72.9 & 4.879e-22 & lte031.0-5.5-2.0a+0.4.BT-Settl.spec.7.dat \cr
  \hline
  \label{tab_MplusTeff_sdM:table_best_fits_FL}
\end{tabular}
\end{table*}
%

%
%%%%% LF option %%%%%
%
\begin{table*}
\centering
\caption{Derived physical parameters for sdM, esdM, and usdM from the comparison between the observed 
X-shooter spectra and the BT-Settl synthetic spectra using the ``LF'' procedure. For each spectral subtype 
and metal class, we list the best fits with effective temperature (T$_{\rm eff}$), gravity ($\log$g), 
metallicity (M/H), chi$^{2}$ value, factor, and the name of the model.
}
\begin{tabular}{c c c c c c c c}
  \hline
  \hline
SpT  & Range  & T$_{\rm eff}$ & $\log$g  & M/H & chi$^{2}$  & Factor & Model \\
  \hline
sdM0.0 & LF & 3700 & 5.5 & -2.5 & 3.692 & 2.021e-22 & lte037.0-5.5-2.5a+0.4.BT-Settl.spec.7.dat \cr
sdM0.5 & LF & 3600 & 4.5 & -1.5 & 87.73 & 1.999e-21 & lte036.0-4.5-1.5a+0.4.BT-Settl.spec.7.dat \cr
sdM1.0 & LF & 3700 & 5.0 & -0.5 & 67.22 & 1.609e-21 & lte037.0-5.0-0.5a+0.2.BT-Settl.spec.7.dat \cr
sdM1.5 & LF & 3500 & 5.0 & -0.5 & 93.49 & 3.502e-22 & lte035.0-5.0-0.5a+0.2.BT-Settl.spec.7.dat \cr
sdM2.0 & LF & 3500 & 4.5 & -1.5 & 60.84 & 1.051e-21 & lte035.0-4.5-1.5a+0.4.BT-Settl.spec.7.dat \cr
sdM2.5 & LF & 3600 & 5.0 & -1.0 & 37.11 & 7.791e-22 & lte036.0-5.0-1.0a+0.4.BT-Settl.spec.7.dat \cr
sdM3.0 & LF & 3500 & 5.0 & -0.0 & 88.51 & 2.208e-21 & lte035.0-5.0-0.0a+0.0.BT-Settl.spec.7.dat \cr
sdM3.5 & LF & 3400 & 5.0 & -1.0 & 169.9 & 3.924e-21 & lte034.0-5.0-1.0a+0.4.BT-Settl.spec.7.dat \cr
sdM4.0 & LF & 3400 & 5.0 & -1.0 & 58.22 & 1.54e-21 & lte034.0-5.0-1.0a+0.4.BT-Settl.spec.7.dat \cr
sdM4.5 & LF & 3300 & 5.0 & -1.5 & 66.55 & 2.115e-22 & lte033.0-5.0-1.5a+0.4.BT-Settl.spec.7.dat \cr
sdM5.0 & LF & 3100 & 5.5 & -0.5 & 98.69 & 6.395e-22 & lte031.0-5.5-0.5a+0.2.BT-Settl.spec.7.dat \cr
sdM5.5 & LF & 3100 & 5.0 & -1.5 & 66.45 & 3.139e-22 & lte031.0-5.0-1.5a+0.4.BT-Settl.spec.7.dat \cr
sdM6.0 & LF & 3100 & 4.5 & -1.5 & 66.98 & 3.779e-22 & lte031.0-4.5-1.5a+0.4.BT-Settl.spec.7.dat \cr
sdM6.5 & LF & 2800 & 4.5 & -2.5 & 8854 & 5.129e-21 & lte028.0-4.5-2.5a+0.4.BT-Settl.spec.7.dat \cr
sdM8.0 & LF & 3000 & 5.0 & -2.5 & 83.22 & 2.962e-22 & lte030.0-5.0-2.5a+0.4.BT-Settl.spec.7.dat \cr
sdM9.5 & LF & 3000 & 5.0 & -2.5 & 63.45 & 8.95e-22 & lte030.0-5.0-2.5a+0.4.BT-Settl.spec.7.dat \cr
  \hline
esdM0.0 & LF & 3800 & 5.0 & -1.0 & 46.36 & 1.789e-22 & lte038.0-5.0-1.0a+0.4.BT-Settl.spec.7.dat \cr                 
esdM0.5 & LF & 3700 & 5.0 & -2.0 & 29.55 & 4.919e-22 & lte037.0-5.0-2.0a+0.4.BT-Settl.spec.7.dat \cr                 
esdM1.0 & LF & 3600 & 4.5 & -2.0 & 72.12 & 2.124e-22 & lte036.0-4.5-2.0a+0.4.BT-Settl.spec.7.dat \cr
esdM1.5 & LF & 3500 & 4.5 & -2.0 & 29.95 & 7.937e-22 & lte035.0-4.5-2.0a+0.4.BT-Settl.spec.7.dat \cr
esdM2.0 & LF & 3500 & 4.5 & -2.0 & 91.63 & 2.571e-22 & lte035.0-4.5-2.0a+0.4.BT-Settl.spec.7.dat \cr
esdM3.0 & LF & 3500 & 5.0 & -1.5 & 97.21 & 6.376e-22 & lte035.0-5.0-1.5a+0.4.BT-Settl.spec.7.dat \cr
esdM3.5 & LF & 3400 & 5.0 & -2.0 & 44.81 & 2.134e-22 & lte034.0-5.0-2.0a+0.4.BT-Settl.spec.7.dat \cr
esdM4.0 & LF & 3400 & 5.0 & -1.5 & 55 & 2.057e-22 & lte034.0-5.0-1.5a+0.4.BT-Settl.spec.7.dat \cr
esdM4.5 & LF & 3200 & 4.5 & -2.0 & 439 & 4.69e-21 & lte032.0-4.5-2.0a+0.4.BT-Settl.spec.7.dat \cr
esdM5.0 & LF & 3200 & 4.5 & -2.0 & 43.03 & 1.99e-22 & lte032.0-4.5-2.0a+0.4.BT-Settl.spec.7.dat \cr
esdM5.5 & LF & 3200 & 5.0 & -2.0 & 92.35 & 1.738e-21 & lte032.0-5.0-2.0a+0.4.BT-Settl.spec.7.dat \cr
esdM6.0 & LF & 3200 & 4.5 & -2.0 & 30.38 & 9.303e-23 & lte032.0-4.5-2.0a+0.4.BT-Settl.spec.7.dat \cr
esdM6.5 & LF & 3200 & 5.0 & -2.0 & 88.16 & 2.966e-22 & lte032.0-5.0-2.0a+0.4.BT-Settl.spec.7.dat \cr
esdM7.0 & LF & 3000 & 4.5 & -2.0 & 10.14 & 1.032e-22 & lte030.0-4.5-2.0a+0.4.BT-Settl.spec.7.dat \cr
esdM7.5 & LF & 3000 & 5.0 & -2.0 & 75.73 & 8.239e-22 & lte030.0-5.0-2.0a+0.4.BT-Settl.spec.7.dat \cr
esdM8.5 & LF & 3000 & 5.0 & -2.5 & 50.75 & 4.173e-22 & lte030.0-5.0-2.5a+0.4.BT-Settl.spec.7.dat \cr
  \hline
usdM0.0 & LF & 3600 & 5.5 & -2.5 & 232.8 & 2.315e-22 & lte036.0-5.5-2.5a+0.4.BT-Settl.spec.7.dat \cr
usdM0.5 & LF & 3800 & 5.5 & -2.0 & 36.52 & 7.582e-23 & lte038.0-5.5-2.0a+0.4.BT-Settl.spec.7.dat \cr
usdM1.0 & LF & 3600 & 5.0 & -2.0 & 39.32 & 1.403e-22 & lte036.0-5.0-2.0a+0.4.BT-Settl.spec.7.dat \cr
usdM2.5 & LF & 3500 & 5.0 & -2.0 & 45.49 & 1.938e-22 & lte035.0-5.0-2.0a+0.4.BT-Settl.spec.7.dat \cr
usdM3.0 & LF & 3600 & 5.5 & -2.0 & 43.27 & 1.199e-22 & lte036.0-5.5-2.0a+0.4.BT-Settl.spec.7.dat \cr
usdM4.0 & LF & 3300 & 5.0 & -2.0 & 10.1 & 2.222e-22 & lte033.0-5.0-2.0a+0.4.BT-Settl.spec.7.dat \cr
usdM4.5 & LF & 3400 & 5.0 & -2.0 & 103 & 2.461e-22 & lte034.0-5.0-2.0a+0.4.BT-Settl.spec.7.dat \cr
usdM5.0 & LF & 3500 & 5.5 & -2.0 & 206.1 & 5.591e-22 & lte035.0-5.5-2.0a+0.4.BT-Settl.spec.7.dat \cr
usdM5.5 & LF & 3400 & 5.5 & -2.0 & 13.36 & 4.883e-23 & lte034.0-5.5-2.0a+0.4.BT-Settl.spec.7.dat \cr
usdM6.0 & LF & 3300 & 5.0 & -2.0 & 49.56 & 4.229e-22 & lte033.0-5.0-2.0a+0.4.BT-Settl.spec.7.dat \cr
usdM7.5 & LF & 3100 & 5.0 & -2.0 & 62.7 & 3.727e-22 & lte031.0-5.0-2.0a+0.4.BT-Settl.spec.7.dat \cr
usdM8.5 & LF & 3100 & 5.5 & -2.5 & 49.69 & 4.503e-22 & lte031.0-5.5-2.5a+0.4.BT-Settl.spec.7.dat \cr
  \hline
  \label{tab_MplusTeff_sdM:table_best_fits_LF}
\end{tabular}
\end{table*}
%

%
%%%%% LF option %%%%%
%
\begin{table*}
\centering
\caption{Derived physical parameters for sdM, esdM, and usdM from the comparison between the
optical region (600--1000nm; ''VV'') of the observed X-shooter spectra and the BT-Settl synthetic spectra
For each spectral subtype 
and metal class, we list the best fits with effective temperature (T$_{\rm eff}$), gravity ($\log$g), 
metallicity (M/H), chi$^{2}$ value, factor, and the name of the model.
}
\begin{tabular}{c c c c c c c c}
  \hline
  \hline
SpT  & Range  & T$_{\rm eff}$ & $\log$g  & M/H & chi$^{2}$  & Factor & Model \\
  \hline
sdM0.0 & VV & 3500 & 4.5 & -2.5 & 1.506 & 2.581e-22 & lte035.0-4.5-2.5a+0.4.BT-Settl.spec.7.dat \cr
sdM0.5 & VV & 3700 & 4.5 & -1.0 & 72.9 & 1.816e-21 & lte037.0-4.5-1.0a+0.4.BT-Settl.spec.7.dat \cr
sdM1.0 & VV & 3700 & 5.0 & -0.5 & 66.12 & 1.636e-21 & lte037.0-5.0-0.5a+0.2.BT-Settl.spec.7.dat \cr
sdM1.5 & VV & 3400 & 4.5 & -1.5 & 48.5 & 3.784e-22 & lte034.0-4.5-1.5a+0.4.BT-Settl.spec.7.dat \cr
sdM2.0 & VV & 3500 & 4.5 & -1.0 & 48.26 & 1.082e-21 & lte035.0-4.5-1.0a+0.4.BT-Settl.spec.7.dat \cr
sdM2.5 & VV & 3500 & 4.5 & -1.0 & 35.1 & 9.046e-22 & lte035.0-4.5-1.0a+0.4.BT-Settl.spec.7.dat \cr
sdM3.0 & VV & 3500 & 5.0 & -0.0 & 89.26 & 2.224e-21 & lte035.0-5.0-0.0a+0.0.BT-Settl.spec.7.dat \cr
sdM3.5 & VV & 3400 & 5.0 & -0.5 & 153.6 & 4.033e-21 & lte034.0-5.0-0.5a+0.2.BT-Settl.spec.7.dat \cr
sdM4.0 & VV & 3400 & 5.5 & -0.5 & 47.24 & 1.627e-21 & lte034.0-5.5-0.5a+0.2.BT-Settl.spec.7.dat \cr
sdM4.5 & VV & 3100 & 4.5 & -1.5 & 64.16 & 2.845e-22 & lte031.0-4.5-1.5a+0.4.BT-Settl.spec.7.dat \cr
sdM5.0 & VV & 3200 & 5.5 & -0.0 & 57.52 & 5.62e-22 & lte032.0-5.5-0.0a+0.0.BT-Settl.spec.7.dat \cr
sdM5.5 & VV & 3200 & 5.5 & -1.0 & 69.85 & 2.79e-22 & lte032.0-5.5-1.0a+0.4.BT-Settl.spec.7.dat \cr
sdM6.0 & VV & 3000 & 4.5 & -1.5 & 60.23 & 4.532e-22 & lte030.0-4.5-1.5a+0.4.BT-Settl.spec.7.dat \cr
sdM6.5 & VV & 2900 & 5.5 & -0.0 & 367.4 & 4.09e-21 & lte029.0-5.5-0.0a+0.0.BT-Settl.spec.7.dat \cr
sdM8.0 & VV & 2800 & 5.0 & -2.5 & 55.29 & 3.348e-22 & lte028.0-5.0-2.5a+0.4.BT-Settl.spec.7.dat \cr
sdM9.5 & VV & 2600 & 4.5 & -2.5 & 28.39 & 1.454e-21 & lte026.0-4.5-2.5a+0.4.BT-Settl.spec.7.dat \cr
 \hline
esdM0.0 & VV & 3600 & 4.5 & -1.5 & 51.33 & 2.339e-22 & lte036.0-4.5-1.5a+0.4.BT-Settl.spec.7.dat \cr
esdM0.5 & VV & 3600 & 4.5 & -1.5 & 33.95 & 5.464e-22 & lte036.0-4.5-1.5a+0.4.BT-Settl.spec.7.dat \cr
esdM1.0 & VV & 3600 & 4.5 & -1.5 & 54.39 & 2.181e-22 & lte036.0-4.5-1.5a+0.4.BT-Settl.spec.7.dat \cr
esdM1.5 & VV & 3400 & 4.5 & -1.5 & 31.73 & 9.081e-22 & lte034.0-4.5-1.5a+0.4.BT-Settl.spec.7.dat \cr
esdM2.0 & VV & 3400 & 4.5 & -1.5 & 65.67 & 2.973e-22 & lte034.0-4.5-1.5a+0.4.BT-Settl.spec.7.dat \cr
esdM3.0 & VV & 3400 & 5.0 & -1.5 & 81.18 & 7.466e-22 & lte034.0-5.0-1.5a+0.4.BT-Settl.spec.7.dat \cr
esdM3.5 & VV & 3400 & 5.0 & -1.5 & 43.41 & 2.206e-22 & lte034.0-5.0-1.5a+0.4.BT-Settl.spec.7.dat \cr
esdM4.0 & VV & 3200 & 4.5 & -2.0 & 61.63 & 2.746e-22 & lte032.0-4.5-2.0a+0.4.BT-Settl.spec.7.dat \cr
esdM4.5 & VV & 3200 & 4.5 & -1.5 & 326.9 & 4.852e-21 & lte032.0-4.5-1.5a+0.4.BT-Settl.spec.7.dat \cr
esdM5.0 & VV & 3200 & 4.5 & -1.5 & 45.1 & 1.963e-22 & lte032.0-4.5-1.5a+0.4.BT-Settl.spec.7.dat \cr
esdM5.5 & VV & 3100 & 4.5 & -2.0 & 88.56 & 2.14e-21 & lte031.0-4.5-2.0a+0.4.BT-Settl.spec.7.dat \cr
esdM6.0 & VV & 3200 & 5.0 & -2.0 & 29.11 & 9.46e-23 & lte032.0-5.0-2.0a+0.4.BT-Settl.spec.7.dat \cr
esdM6.5 & VV & 3100 & 4.5 & -2.0 & 86.41 & 3.646e-22 & lte031.0-4.5-2.0a+0.4.BT-Settl.spec.7.dat \cr
esdM7.0 & VV & 3000 & 4.5 & -2.0 & 5.132 & 9.173e-23 & lte030.0-4.5-2.0a+0.4.BT-Settl.spec.7.dat \cr
esdM7.5 & VV & 3000 & 5.0 & -2.0 & 89.47 & 8.325e-22 & lte030.0-5.0-2.0a+0.4.BT-Settl.spec.7.dat \cr
esdM8.5 & VV & 2800 & 4.5 & -2.5 & 52.36 & 5.534e-22 & lte028.0-4.5-2.5a+0.4.BT-Settl.spec.7.dat \cr
 \hline
usdM0.0 & VV & 3500 & 4.5 & -2.5 & 51.01 & 2.583e-22 & lte035.0-4.5-2.5a+0.4.BT-Settl.spec.7.dat \cr
usdM0.5 & VV & 3600 & 5.0 & -2.5 & 42.57 & 9.734e-23 & lte036.0-5.0-2.5a+0.4.BT-Settl.spec.7.dat \cr
usdM1.0 & VV & 3600 & 5.0 & -1.5 & 40.24 & 1.445e-22 & lte036.0-5.0-1.5a+0.4.BT-Settl.spec.7.dat \cr
usdM2.5 & VV & 3400 & 4.5 & -2.0 & 50.77 & 2.315e-22 & lte034.0-4.5-2.0a+0.4.BT-Settl.spec.7.dat \cr
usdM3.0 & VV & 3400 & 4.5 & -2.0 & 40.27 & 1.684e-22 & lte034.0-4.5-2.0a+0.4.BT-Settl.spec.7.dat \cr
usdM4.0 & VV & 3500 & 5.5 & -2.0 & 2.578 & 1.425e-22 & lte035.0-5.5-2.0a+0.4.BT-Settl.spec.7.dat \cr
usdM4.5 & VV & 3300 & 5.0 & -2.0 & 83.97 & 2.923e-22 & lte033.0-5.0-2.0a+0.4.BT-Settl.spec.7.dat \cr
usdM5.0 & VV & 3400 & 5.5 & -2.5 & 131.2 & 6.592e-22 & lte034.0-5.5-2.5a+0.4.BT-Settl.spec.7.dat \cr
usdM5.5 & VV & 3800 & 6.0 & -0.5 & 16.71 & 2.727e-23 & lte038.0-6.0-0.5a+0.2.BT-Settl.spec.7.dat \cr
usdM6.0 & VV & 3200 & 5.0 & -2.0 & 44.45 & 4.976e-22 & lte032.0-5.0-2.0a+0.4.BT-Settl.spec.7.dat \cr
usdM7.5 & VV & 3000 & 5.0 & -2.5 & 28.43 & 4.579e-22 & lte030.0-5.0-2.5a+0.4.BT-Settl.spec.7.dat \cr
usdM8.5 & VV & 3000 & 5.5 & -2.5 & 51.61 & 5.169e-22 & lte030.0-5.5-2.5a+0.4.BT-Settl.spec.7.dat \cr
  \hline
  \label{tab_MplusTeff_sdM:table_best_fits_VV}
\end{tabular}
\end{table*}

\end{appendix}

\end{document}